\documentclass[longbib,PRB,reprint,superscriptaddress]{revtex4-1}
\usepackage{amsmath}
\usepackage{graphicx}
\usepackage{braket}
\usepackage{url}
\usepackage[T1]{fontenc}
\usepackage[utf8]{inputenc}
\usepackage{amssymb}
\usepackage[unicode=true]{hyperref}
\usepackage[usenames,dvipsnames]{color}

\begin{abstract}
We investigate the interaction between the electrons of a two-dimensional metal and the acoustic phonons of an underlying piezoelectric substrate. Fundamental inequalities can be obtained from general energy arguments. As a result, phonon mediated attraction can be proven to never overcome electron Coulomb repulsion, at least for long phonon wavelengths. We study the influence of these phonons on the possible pairing instabilities of a two-dimensional electron gas such as graphene.
\end{abstract}

\makeatletter

\begin{document}

\title{Electron-phonon vertex and its influence on the superconductivity of two-dimensional metals on a piezoelectric substrate}

\author{David G. Gonz\'alez}
\email[email:\,]{d.gonzalez@mat.ucm.es}
\affiliation{Departamento de F\'isica de Materiales, Universidad Complutense de Madrid, E-28040 Madrid, Spain}
\affiliation{Campus de Excelencia Internacional, Campus Moncloa UCM-UPM, E-28049 Madrid, Spain}
\author{Fernando Sols}
\affiliation{Departamento de F\'isica de Materiales, Universidad Complutense de Madrid, E-28040 Madrid, Spain}
\affiliation{Campus de Excelencia Internacional, Campus Moncloa UCM-UPM, E-28049 Madrid, Spain}
\author{Francisco Guinea}
\affiliation{%
IMDEA Nanociencia, Calle de Faraday~9,
E-28\,049 Madrid, Spain}
\affiliation{%
Department of Physics and Astronomy, University of Manchester, Oxford Road, Manchester M13~9PL, United Kingdom
}
\author{Ivar Zapata}
\affiliation{Departamento de F\'isica de Materiales, Universidad Complutense de Madrid, E-28040 Madrid, Spain}

\date{\today}
\maketitle

\section{Introduction}

Surface acoustic waves (SAWs) \cite{Landau1986} have been used for decades as a valuable scientific and technological tool.
In the context of electronics,
they are often excited in piezoelectric materials
\cite{Hutson1962,Morgan2010,Auld1990,Royer2000,Royer2000a,Newnham2004},
where the mechanical and electrical fields are coupled. In particular, they have been applied as experimental probes of the quantum Hall effects in
two-dimensional electron gases (2DEG) \cite{Wixforth1986}.

On the other hand, there is an increased interest in 2DEG since the isolation of graphene in 2004 and the production of other two-dimensional (2D) materials which followed it.
Due to their unusual character, the properties of graphene electrons have been intensively studied during the last decade
\cite{CastroNeto2009}. Although graphene on a substrate has received considerable attention, relatively few studies have been devoted to the case of graphene in contact with a piezoelectric material.
These include the propagation of surface acoustic waves on graphene \cite{Thalmeier2010}, some acoustoelectric effects \cite{Miseikis2012} \cite{Bandhu2013}, and the relaxation induced by the
surface acoustic wave quanta on graphene electrons \cite{Zhang2013}, among others.
Recently, it has been proposed that surface acoustic waves (SAW) can provide a
diffraction grating for the conversion of light into graphene plasmons \cite{Schiefele2013}.

The coupling of the piezoelectric SAW to the electrons in
a 2DEG or in graphene has been computed, within certain simplifying assumptions and for definite substrate crystal structures, in Ref. \cite{Simon1996,Knabchen1996}. The derivation of the electron-surface phonon interaction for a piezoelectric material has been performed only within a purely elastic Rayleigh wave approximation \cite{Knabchen1996} or for definite propagating directions \cite{Simon1996}. But these methods fail in stronger piezoelectric materials and for other crystal symmetries. For instance, the isotropic Rayleigh wave approximation in the case of lithium niobate leads to surface acoustic wave velocities about 15 \% too low and lacking the correct angular dependence \cite{Morgan2010,Farnell1970}, and this material is not the one with the largest electromechanical couplings at all. Moreover, the obtained vertex are expressed in terms of a matching constant whose physical interpretation is rather obscure, allowing for just order of magnitude estimates.

In the present work we calculate a general electron-phonon interaction [see Eq. (\ref{eqn:PAVertex})], which is expressed solely in terms of physical quantities characterizing the response of the substrate surface. We emphasize that all the quantities appearing in the vertex are both computable from linear piezo-elasticity theory and experimentally measurable. One of them, the {\it electromechanical coupling coefficient}, $K_R$, will turn out to be central to all computations, serving as a natural dimensionless parameter which provides the scale for the effect of the substrate piezoelectricity on the 2D electron system. Moreover, from very general considerations explained in Appendix \ref{app:PiezoElectricVertex}, we are able to provide bounds on its size: $0\le K_R<1$ [see Eq. (\ref{eqn:interactionsRatio})]. It is important to note that the vertex written here is derived within the framework of linear piezo-elastic theory, which means that its validity should be restricted to low amplitude, low frequency and long wavelength phenomena. Bulk modes are also left aside in this work. On the other hand, our study is not restricted to any approximation based on the symmetry or the piezoelectric softness of the substrate.

Equipped with the effective electron-electron interaction which results from taking into account the exchange of these acoustic phonons between the electrons in graphene (or other 2D materials), the question can be raised of whether these interactions might be attractive and, depending on some material parameters and the tunable electronic density of graphene, perhaps strong enough to generate electron Cooper pairing and superconductivity \cite{Mahan2013}. We can further ask whether such a superconductivity could be observed at temperatures attainable in a laboratory without the recurring to huge non gate-achievable doping levels, as predicted for intrinsic graphene phonons \cite{Einenkel2011} or for Kohn-Luttinger or electronic superconductivity in other graphene heterostructures with repulsive interactions \cite{Guinea2012}. From the interaction vertex derived in the present work, it can be shown that the relative size of the static phonon-mediated electron-electron interaction with respect to the original Coulomb repulsion turns out to be exactly $K_R^2<1$, because of the aforementioned general inequality. However, by applying the Eliashberg formalism to graphene \cite{Einenkel2011}, we are able to assess, in terms of $K_R$, the influence that these low-frequency and long-wavelength phonons have on possible BCS type instabilities. The conclusion is that  present piezoelectric materials are not able to either induce $s-$wave pairing by themselves or affect in a significant way any pairing instability which could be already present in graphene. We note, however, that this conclusion could be substantially changed in case new hard piezoelectric materials would be found.

\section{Electron-phonon interaction}
\begin{figure}
	\centering
	\includegraphics[width=0.9\linewidth]{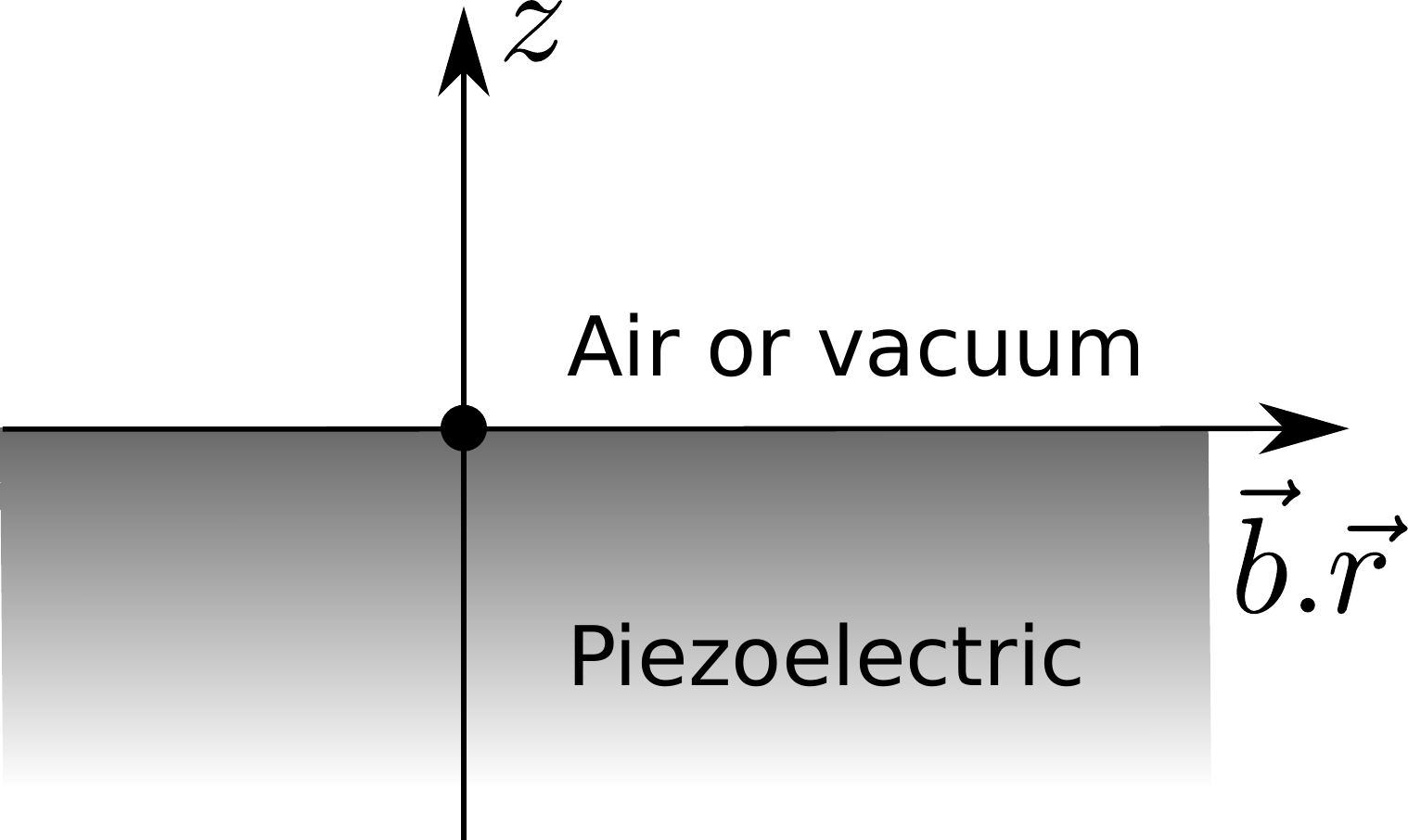}
	\caption{A flat piezoelectric substrate with an interface to air or vacuum, with $\mathbf{b}=\left[\cos(\theta),\sin(\theta)\right]$ the propagation direction of the piezoelectric SAW.}
	\label{fig:PiezoSAWSchema}
\end{figure}

The system to be considered is depicted in Fig.~\ref{fig:PiezoSAWSchema}. The $z=0$ interfacial surface is supposed to be free of tension and, when acting as a substrate to a deposited graphene sheet (or any other 2D charged system), free of electrodes as well. However, by introducing a surface charge, one can express the 2D response of the piezoelectric substrate, the {\it piezoelectric surface permittivity}, as the ratio of an electric displacement to an electric field. To be precise, let us allow for a surface charge with a harmonic dependence along $\textbf{q}=q[\cos \theta,\sin\theta]$ (here $\textbf{R}=(x,y)$)
\begin{align}
	\sigma(\mathbf{R},t)=\sigma(\mathbf{q},\omega) e^{i (\textbf{q}\cdot\textbf{R}-\omega t)},
\label{eqn:surfaceChargeDef}
\end{align}
and, from linearity, all quantities evaluated at the surface have the same 2D space-time dependence. We use SI units throughout this work. Because the medium at $z>0$ has a dielectric constant $\varepsilon_\text{vac}$, then, from Poisson's equation it follows that
\begin{align}
	\varepsilon_{\text{vac}}= \frac{D_3^+(\mathbf{q},\omega)}{q \varphi(\mathbf{q},\omega)},
\label{eqn:vaccumPermittivity}
\end{align}
where $D_3^+$ is the normal to the surface Fourier component of the electric displacement field over the surface (on the $z \rightarrow 0^+$ side) and $\varphi$ is the electric potential, which is continuous because we do not allow for anything more singular than surface charges. The same ratio taken below the surface ($z \rightarrow 0^-$ ) allows us to introduce the (relative) {\it piezoelectric surface permittivity},
\begin{align}
	\tilde{\varepsilon}(\mathbf{q},\omega) := -\frac{1}{\varepsilon_{\text{vac}}}\frac{D_3^-(\mathbf{q},\omega)}{q \varphi(\mathbf{q},\omega)} \, ,
\label{eqn:piezoSurfacePermittivityDef}
\end{align}
which can also be straightforwardly expressed in terms of the surface impedance tensor \cite{Ingebrigsten1969}.
From Poisson's equation and Eqs. (\ref{eqn:surfaceChargeDef}-\ref{eqn:piezoSurfacePermittivityDef}) it follows
\begin{equation}\label{eqn:surfaceChargePotentialRelation}
	\sigma(\mathbf{q},\omega)=D_3^+(\mathbf{q},\omega)-D_3^-(\mathbf{q},\omega)=q \varphi(\mathbf{q},\omega) [1+\tilde{\varepsilon}(\textbf{q},\omega)]\varepsilon_\text{vac}.
\end{equation}

Further analysis summarized in Appendix \ref{app:PiezoElectricVertex} shows that $\tilde{\varepsilon}(\mathbf{q},\omega)$ has a dependence of the form
$\tilde{\varepsilon}(\mathbf{q},\omega)=\tilde{\varepsilon}(\mathbf{q}/\omega)$. An immediate conclusion from Eq. (\ref{eqn:surfaceChargePotentialRelation}) is that a purely piezoelectric wave (i.e., without sources, $\sigma(\mathbf{q},\omega)=0$), can propagate without damping if and only if $\tilde{\varepsilon}(\mathbf{q}/\omega)+1=0$. Thus, if the phase velocity is $v_s(\theta)$, then $\tilde{\varepsilon}(\mathbf{b}/v_s(\theta))+1=0$ and the dispersion relation of the obtained, called piezoelectric Rayleigh waves (here referred to as SAW) is $\omega=v_s(\theta) q$.

For the high-frequency limit we introduce $\tilde{\varepsilon}_{\text{HF}}(\theta):=\tilde{\varepsilon}(\mathbf{q}/\omega), \omega\rightarrow\infty$. This should be the anisotropic dielectric function valid all the way up to the optical region, and should take into account all screening processes in the substrate except for the slow piezoelectric ones, which are estimated below for the substrate of the 2D electronic material.

A central quantity in the evaluation of devices which use piezoelectric Rayleigh waves is the {\it SAW electromechanical coupling coefficient} $K_R(\theta)$, introduced through the relation at $1+\tilde{\varepsilon}(\textbf{b}/v_s(\theta))=0$:
\begin{equation}
	\frac{K_R^2(\theta)/2}{\tilde{\varepsilon}_{\text{HF}}(\theta)+1} = \left[v_s(\theta)\frac{\partial \tilde{\varepsilon}(\textbf{b}/v)}{\partial v}\Bigr|_{v=v_s(\theta)} \right]^{-1} ~.
\label{eqn:emCouplingCoefficient}
\end{equation}

In Appendix \ref{app:PiezoElectricVertex} we show that very general considerations require
\begin{equation} \label{eqn:central-inequality}
0 \le K_R(\theta)<1\, , 
\end{equation}
which is one of the central results of this work. The inequality $(\ref{eqn:central-inequality})$  is crucial because, as we will show, it implies that at small frequencies piezoelectric phonons cannot provide the sufficient screening to overcome the bare Coulomb repulsion.

In Ref. \cite{Royer2000} it is shown that there is a relation between the amplitude of electric potential at the surface, $\varphi_0=\varphi(\mathbf{q},\omega)$ and the total energy $H_{\text{harm}}$, see Eq. (\ref{eqn:HamiltonianHarmonicPSAW}). Hence, standard quantization procedure (see the Appendix, subsection \ref{app:HamiltonianInterVertex}) shows that the interaction between the 2D electronic material sheet and the spontaneous piezoelectric Rayleigh waves can be written as
\begin{align}
	H_{\text{e-ph}}^{\text{PA}} &= \frac{1}{\sqrt{A}} {\sum_{\mathbf{k,q},\sigma}}\gamma_{\mathbf{q}}^{\text{PA}} \, a_{\mathbf{k+q},\sigma}^{\dagger}a_{\mathbf{k},\sigma}b_{\mathbf{q}}+\text{H.c.}~,
	\nonumber \\
	\gamma^{\text{PA}}_\mathbf{q} & = \frac{K_R(\theta)}{2} \sqrt{\frac{ \hbar e^2 v_s(\theta)}{\overline{\varepsilon}_{\text{HF}}(\theta)\varepsilon_\text{vac}}}=K_R(\theta)\sqrt{\frac{\pi\alpha_\text{fs}\hbar^2 v_F v_s(\theta)}{\overline{\varepsilon}_\text{HF}(\theta)}},
\label{eqn:PAVertex}
\end{align}
where $A$ is the area of the sample, $a_{\mathbf{k},\sigma},a^\dagger_{\mathbf{k},\sigma}$ the electron operators with $\sigma=\pm$ the electron spin, $b_{\mathbf{k}},b^\dagger_{\mathbf{k}}$ are the piezoelectric phonon operators, $\alpha_\text{fs}:=e^2/4\pi\varepsilon_\text{vac}\hbar v_F$ and  $\overline{\varepsilon}_{\text{HF}}(\theta) := (\tilde{\varepsilon}_{\text{HF}}(\theta)+1)/2$. The validity of the shown interaction hamiltonian Eq. (\ref{eqn:PAVertex}) requires two further assumptions: first, the 2DEG or multilayered graphene sample should be thin enough so that in Eq. (\ref{eqn:PSAWWaveVacuum}), $kd\ll 1$, where $z\equiv d$ is the width of the sample and $k$ is the maximum allowed phonon momentum. And second, this maximum allowed momentum should be sufficiently small for the classical piezo-elasticity theory, as shown in Eqs. (\ref{eqn:elasticNewton}-\ref{eqn:piezoConstitutive}), to remain valid. We assume that a maximum momentum on the order of $k_F \sim 10^6-10^7 \, \text{cm}^{-1}$ does not violate this last restriction.

The resulting total Hamiltonian for the combined system of 2D electron gas and piezoelectric Rayleigh phonon is
\begin{eqnarray}
	H= \sum_{\mathbf{k},\sigma}E_{\mathbf{k}} a_{\mathbf{k},\sigma}^{\dagger} a_{\mathbf{k},\sigma} + \hbar \sum_{\mathbf{q}}\omega_{\mathbf{q}}\, b_{\mathbf{q}}^{\dagger} b_{\mathbf{q}} \nonumber \\
	+  H_{\text{e-ph}}^{\text{PA}} + \frac{1}{2A} \sum_{\bf{q}} v_{\mathbf{q}}^{(0)} \rho(\mathbf{q}) \rho(\mathbf{-q} )~,
\label{eqn:PAHamiltonian}
\end{eqnarray}
where $E_\mathbf{k}$ is the electron energy for a 2D wave-vector $\mathbf{k}$,
$\omega_\mathbf{q}=v_s(\theta) q$ is the dispersion relation for the acoustic piezoelectric SAW phonon of 2D wave-vector $\mathbf{q}$ and $v_s(\theta)$ is the SAW propagation velocity, and
$\rho(\mathbf{q}) = \sum_{\mathbf{k},\sigma} a_{\mathbf{k+q},\sigma}^{\dagger} a_{\mathbf{k},\sigma}$ is the Fourier transform of the electron density.

We use the bare Coulomb electron-electron interaction as
\begin{align}
	v_{\mathbf{q}}^{(0)} &= \frac{ e^{2}}{2\overline{\varepsilon}_{\text{HF}}(\theta) \varepsilon_\text{vac} q},
\end{align}
which contains all high-frequency screening processes except for piezoelectric ones.

The bare electron-electron interaction mediated by phonons is \cite{Hwang2010,Mahan2013}
\begin{equation}
	V_{\text{ph}}^{\text{PA}}(\mathbf{q},\omega) = |\gamma_{\mathbf{q}}^{\text{PA}}|^{2} G_{0}^{\text{PA}}(\mathbf{q},\omega) \, ,
%= 2|\gamma_{\mathbf{q}}^{\text{PA}}|^{2} \frac{\hbar v_{s}q}{\hbar^{2}\omega^{2}-\hbar^{2}v_{s}^{2}q^{2}}
	\label{eqn:bareElectronPhononElectronInteraction}
\end{equation}
where
\begin{equation}
G_{0}^{\text{PA}}(\mathbf{q},\omega) = \frac{2\omega_\mathbf{q}}{\hbar\left(\omega^2-\omega_\mathbf{q}^2+i\eta\right)}.
\end{equation}
is the bare piezoelectric acoustic phonon propagator ($\eta \rightarrow 0^+$).  The resulting RPA-type approximation to the dielectric function and effective interaction are:
\begin{align}
	V_{\text{eff}}(\mathbf{q},\omega) &=\frac{ e^2}{2\varepsilon(\mathbf{q},\omega)\varepsilon_\text{vac}q} \nonumber \\
	 &=\frac{v_{\mathbf{q}}^{(0)}+V_{\text{ph}}^{\text{PA}}(\mathbf{q},\omega)}{1-\left[v_{\mathbf{q}}^{(0)}+V_{\text{ph}}^{\text{PA}}(\mathbf{q},\omega)\right]\,\Pi_{0}(\mathbf{q},\omega)} %\nonumber 
	\label{eqn:V_eff}
\end{align}
which can also be written as
\begin{equation} \label{eqn:V-eff-RPA}
V_{\text{eff}}(\mathbf{q},\omega)
= \frac{v_{\mathbf{q}}^{(0)}}{\varepsilon_{\text{RPA}}(\mathbf{q},\omega)} + \left|\frac{\gamma_{\mathbf{q}}^{\text{PA}}}{\varepsilon_{\text{RPA}}(\mathbf{q},\omega)} \right|^{2} \tilde{G}^{\text{PA}}(\mathbf{q},\omega) ~.
\end{equation}
where $\varepsilon_{\text{RPA}}(\mathbf{q},\omega) \equiv 1-v_{\mathbf{q}}^{(0)}\Pi_{0}$, with  $\Pi_{0}$ the irreducible polarization function,
\begin{equation}
\tilde{G}^{\text{PA}}(\mathbf{q},\omega) =
\frac{G_{0}^{\text{PA}}(\mathbf{q},\omega)}
{1-\frac{|\gamma_{\mathbf{q}}^{\text{PA}}|^{2}G_{0}^{\text{PA}}(\mathbf{q},\omega)\Pi_{0}(\mathbf{q},\omega)}{\varepsilon_{\text{RPA}}(\mathbf{q},\omega)}}\, .
\end{equation}
In the low frequency limit, $\Pi_{0}(\mathbf{q},\omega\simeq 0)\simeq -D(E_F)=-2k_F/\pi \hbar v_F$, for monolayer graphene \cite{Wunsch2006}, or $\Pi_{0}(\mathbf{q},\omega\simeq 0)\simeq -D(E_F)=-m/2\pi \hbar^2$ for a 2DEG with effective mass $m$ \cite{Giuliani2005}. These two last static limits are exact for $q<2k_F$. In Eq. (\ref{eqn:V-eff-RPA}), the total interaction has been rewritten as the sum of a purely electronically screened Coulomb repulsion and a phonon-induced effective part in which the vertex and phonon-propagator are also screened by just the conducting electrons of the 2DEG or graphene \cite{Mahan2013,Mattuck2012}.
%\begin{align}
%\varepsilon_{\text{RPA}}(\mathbf{q},\omega) &= 1- v_{\mathbf{q}}^{(0)}(\mathbf{q})\Pi_{0}(\mathbf{q},\omega)~, \label{eqn:dielectricFunctionRPA} \\
%\tilde{G}^{\text{PA}}(\mathbf{q},\omega)&= \frac{G_{0}^{\text{PA}}(\mathbf{q},\omega)}{1-\frac{|\gamma_{\mathbf{q}}^{\text{PA}}|^{2}G_{0}^{\text{PA}}(\mathbf{q},\omega)\Pi_{0}(\mathbf{q},\omega)}{\varepsilon_{\text{RPA}}(\mathbf{q},\omega)}}~. \label{eqn:dressedPhononPropagator}
%\end{align}
For frequencies small in the scale of the acoustic phonons (or the Bloch-Gr\"uneisen temperature $k_B T_{\rm BG}:=2\hbar v_s k_F $), the bare electron-phonon-electron interaction contributes to the long-range part of the total interaction with a ${\bf q}$-dependence similar to that of the Coulomb repulsion:
\begin{equation}
	V_{\text{ph}}^{\text{PA}}(\mathbf{q},\omega \simeq 0)=|\gamma_{\mathbf{q}}^{\text{PA}}|^{2}G_{0}^{\text{PA}}(\mathbf{q},\omega \simeq 0) = - \frac{2|\gamma_{\mathbf{q}}^{\text{PA}}|^{2}}{\hbar v_{s}q}~,
\label{eqn:Vstatic}
\end{equation}
 Note that it is the acoustic phonon propagator $G_{0}^{\text{PA}}(\mathbf{q},\omega \simeq 0)$ that introduces the coulombic long-range dependence in $q$ via the dispersion of the modes. In the next subsection we shall see that a similar final $q$ dependence has a different origin.

In the limit of low frequencies, $\omega \simeq 0$,  there can be no effective attraction for electrons close to the Fermi-surface because, as shown in Eq. (\ref{eqn:kSquareBound}), the following inequality is satisfied
\begin{equation}
	\frac{-V_{\text{ph}}^{\text{PA}}(\mathbf{q},\omega \simeq 0)}{v_{\mathbf{q}}^{(0)}}=K_R^2(\theta)<1 \label{eqn:interactionsRatio} \,
\end{equation}
which is, in conjunction with the interaction vertex given by Eq. (\ref{eqn:PAVertex}), a central result of this paper.

\subsection{Comparison with optical phonons}

For simplicity, we focus a single branch of the longitudinal optical (LO) for which we assume a constant frequency $\omega_0$. The total Hamiltonian reads as in Eq. (\ref{eqn:PAHamiltonian}) except for the replacements:
\begin{align}
\omega_{\mathbf{q}} & \rightarrow \omega_0 \\
v^{(0)}_{\mathbf{q}} & \rightarrow v^{(\infty)}_{\mathbf{q}} := \frac{e^2}{2\overline{\varepsilon}_\infty \varepsilon_\text{vac} q} \\
\overline{\varepsilon}_\infty & := \frac{\varepsilon_\infty+1}{2} \\
\gamma^{\text{PA}}_\mathbf{q} & \rightarrow \gamma^{\text{OP}}_\mathbf{q} := \sqrt{ g \frac{e^ 2 \hbar \omega_0}{2\varepsilon_\text{vac}q}} \\
g & := \left(\frac{1}{\varepsilon_\infty+1} - \frac{1}{\varepsilon_0+1}\right) >0\, ,
\end{align}
where standard notation for dielectrics is used: $\varepsilon_\infty$ the dielectric constant coming from very high frequency interband electronic transitions and $\varepsilon_0$ would be static dielectric constant in the absence of the piezoelectric phonons at frequencies much smaller than $\omega_0$. 
%Here we are assuming that the cut on the substrate is such that no 2D inhomogeneity terms appear. The modification to include them is straightforward.
	
Again, as shown in the discussion around Eq. (\ref{eqn:Vstatic}), for small frequencies ($\omega \ll \omega_0$), the bare phonon-mediated electron-electron interaction contributes to the long-range part of the total interaction like the Coulomb repulsion:
\begin{equation}
V_{\text{ph}}^{\text{OP}}(\mathbf{q},\omega \simeq 0)=|\gamma_{\mathbf{q}}^{\text{OP}}|^{2}G_{0}^{\text{OP}}(\mathbf{q},\omega \simeq 0) = - g \frac{e^2}{\varepsilon_\text{vac} q}~. \label{eqn:VOPstatic}
\end{equation}
However, in contrast to the piezoelectric case, here it is the vertex that introduces the coulombic dependence in $q$.

At small frequencies, $\omega \ll \omega_0$, a single optical phonon is not enough to provide over-screening, because
\begin{equation}
	\frac{-V_{\text{ph}}^{\text{OP}}(\mathbf{q},\omega \simeq 0)}{v_{\mathbf{q}}^{(\infty)}}= \frac{\varepsilon_0-\varepsilon_\infty}{\varepsilon_0+1}<1~.
	\label{eqn:optInteractionsRatio}
\end{equation}

\section{Effect of piezoelectric phonons on superconducting instabilities}

From (\ref{eqn:interactionsRatio}) and (\ref{eqn:V_eff}), we see that, in the static limit ($\omega \simeq 0$), and for $q<2k_F$,
$V_{\text{eff}}$ can be written in the form
\begin{equation}
V_{\text{eff}} (\mathbf{q},0) = \frac{[1-K_R^2(\theta)]v_{\mathbf{q}}^{(0)}}{1+\left[1-K_R^2(\theta)\right]v_{\mathbf{q}}^{(0)}D(E_F)}~, 
%\; \; \; \; \; \; (\text{piezo}+2\text{D})
\label{eqn:Pairing2dpiezo}
\end{equation}
where we note that we have not assumed $q\ll k_F$, as discussed in the paragraph following (\ref{eqn:PAVertex}). From the inequality in (\ref{eqn:interactionsRatio}), we are led to conclude 
that over-screening of the Coulomb repulsion by the phonon-mediated attraction is not possible.
Moreover, and following standard textbook reasoning (see for example \cite{Ashcroft2011}), we conclude that BCS-type instabilities must also be ruled out. 
A similar result holds for a single branch of optical phonons, as can be seen from Eq. (\ref{eqn:optInteractionsRatio}) (see however \cite{Gorkov2015} for the effect of multiple optical phonon branches from the substrate on superconducting instabilities). 

Moreover, in case such over-screening occurred, the static dielectric constant from Eq. (\ref{eqn:V_eff}) would predict unphysical features such as unstable phononic modes with $\tilde{\omega}(q_{c}) = 0$ for some $q_{c} \neq 0$ and even imaginary frequencies for $q<q_{c}$. No matter how small the absolute difference $|1-K_R^2(\theta)|$ happened to be, there would always exist a pole for the static (\ref{eqn:V_eff}) at small enough $q$ (what cannot occur in standard BCS metals), signaling a different type of instability, possibly a charge density wave.

On the other hand, the result (\ref{eqn:interactionsRatio}) for the vertex could still lead to higher angular momentum pairing instabilities (as in the Kohn-Luttinger mechanism \cite{kohn1965new}) provided that $K_R^2(\theta)$ is sufficiently large and anisotropic, a case not considered by us.

\subsection{Eliashberg formalism \cite{Einenkel2011}}

The previous reasoning about the absence of superconducting instabilities, is incomplete and somewhat oversimplified. Three reasons support this claim: (i) Long-wave piezoelectric phonon excitations (as considered in the present work) can never be the only source of effective electron-electron interactions; in particular, we have not taken into account the short range electric fluctuations of the substrate. (ii) There is definitely some dynamic over-screening at high frequencies [see Eq. (\ref{eqn:bareElectronPhononElectronInteraction})]. And (iii) Coulomb interaction has to be properly renormalized by taking into account collisions with high momentum transfer, which diminishes the Coulomb repulsion and thus comparatively strengthens the other attraction mechanisms.

Leaving aside the first objection momentarily, we can use the Eliashberg formalism, as applied to graphene in Ref. \cite{Einenkel2011}, to deal with the other two objections. The effective interaction could cause superconducting instabilities if a dimensionless electron-phonon coupling $\lambda^\text{PA}$ happened to be greater than an also dimensionless Coulomb pseudopotential $\mu^*$ coming from high-energy renormalizations \cite{Morel1962,Einenkel2011}. The coupling constant $\lambda^\text{PA}$ in the Eliashberg formalism is the same appearing in (the real part of the) self-energy calculations to renormalize the Fermi velocity \cite{Gonzalez2015} and is given by:
\begin{align}
&\alpha_\text{PA}^2F(\omega)=\frac{|\gamma^\text{PA}|^{2}}{2 \pi^2 \hbar^2 v_s v_F}\frac{\sqrt{1-(\frac{\omega/v_s}{2k_F})^2}}{(1+\frac{k_\text{TF}}{\omega/v_s})^2}~, \nonumber \\
&\lambda^\text{PA}=2\int\limits_{0}^{\infty}\frac{\alpha_\text{PA}^2F(\omega)}{\omega}d\omega = \frac{r_s}{\pi} K_R^2\, F(2r_s)~, \label{eqn:lambda}\\
& F(x) = \int\limits_{0}^{1}\frac{t\sqrt{1-t^2}\,dt}{(t+x)^2}=-2+x\pi + \frac{(1-2x^2)\, \text{acosh}(x^{-1})}{\sqrt{1-x^2}}~, \nonumber
\end{align}
where $r_s(\theta):=\alpha_{\text{fs}}/\varepsilon(\theta)$, and the symbols $r_s, v_s, \gamma^{\text{PA}}, K_R^2$ stand for the Fermi surface angle-averaged quantities of the same name. The constant $\mu^*$ equals
\begin{equation}
\mu^* = \frac{\frac{1}{4}D(E_F)V}{1+\frac{1}{4}D(E_F)V \log(\frac{E_F}{\hbar \omega_c})}~,
\label{eqn:PseudoPotential}
\end{equation}
where $\omega_c$ is some energy cutoff which should satisfy $\omega_\text{Debye} \ll \omega_c \ll E_F/\hbar $ \cite{Einenkel2011} and $V$ comes from the Fermi surface average of the Thomas-Fermi renormalized Coulomb repulsion $v_\mathbf{q}^{(0)}/(1+\frac{k_\text{TF}}{q})$. We have
\begin{align}
&\frac{1}{4}D(E_F)V=\frac{r_s}{\pi}G(2r_s), \\
& G(x)=\int\limits_{0}^{1}\frac{\sqrt{1-t^2}\,dt}{t+x}=-1+\frac{\pi x}{2}+\sqrt{1-x^2}\,\text{acosh}(x^{-1})~, \nonumber
\end{align}
and therefore, provided that one takes $\hbar\omega_c \simeq k_B T_\text{BG}$, so that $\log\left(\frac{E_F}{\hbar \omega_c}\right) \simeq \log(\frac{v_F}{2v_s}) \simeq 5$. Thus, an estimate of the effective pseudo-potential is
\begin{equation}
\mu^*\simeq\frac{\frac{r_s}{\pi}G(2r_s)}{1+\frac{5r_s}{\pi}G(2r_s)} ~.
\end{equation}
There could be intravalley \footnote{In this analysis, we are taking into account only long-wavelength piezoelectric phonons, hence no intervalley pairing instability could occur. We address this question in the next paragraph.} superconducting instabilities provided that
\begin{equation}
1<\frac{\lambda^\text{PA}}{\mu^*}=K_R^2 \frac{F(2r_s)}{G(2r_s)}\left[1+\frac{5r_s}{\pi}G(2r_s)\right]~,
\end{equation}
which imposes a constraint on the value of $K_R^2$ from the piezoelectric substrate with respect to quantities depending on $r_s$. The coupling $K_R$ should be very large and actually greater than 1 for this choice of $\omega_c$, although there could exist superconductivity in this idealized case of a system consisting just of the graphene electrons and long wavelength piezoelectric phonons, provided that $E_F/\omega_c$ is larger and $K_R^2$ close to 1.

In order to amend the first objection, we have to consider proper phonons of the electronic system (here we go on considering graphene), in conjunction with the short range of the piezoelectric ones. Then, pairing instabilities due to intervalley scattering have to be considered as well, because intravalley scattering terms contribute also to the intervalley pairing gap.
With the notation in Ref. \cite{Einenkel2011}, an estimate on the critical temperature for the intravalley pairing is \cite{Einenkel2011} $T_c^\text{intra} = 1.13\omega_\text{Debye}\,\exp\left(-\frac{1+\lambda}{\lambda_{11}-\mu_{11}^*}\right)$, and a very similar is obtained for the intervalley transition
$T_c^{\text{inter}} = 1.13\omega_\text{Debye}\,\exp\left(-\frac{1+\lambda}{\lambda-\mu_{12}^*}\right)$, with $\lambda = \lambda_{11}+\lambda_{12}$ and the previously computed $\lambda^{\text{PA}}$ included into the intravalley term $\lambda_{11}$ ($\lambda_{12}$ denotes the contribution from all intervalley terms). Here the pseudo-potential $\mu_{12}^*$ is only slightly larger than $\mu_{11}$, and both are given by similar formulas as in Eq. (\ref{eqn:PseudoPotential}), but with $\omega_c\rightarrow \omega_\text{Debye}$. 

\begin{figure}
\raggedleft
\includegraphics[width=0.92\linewidth]{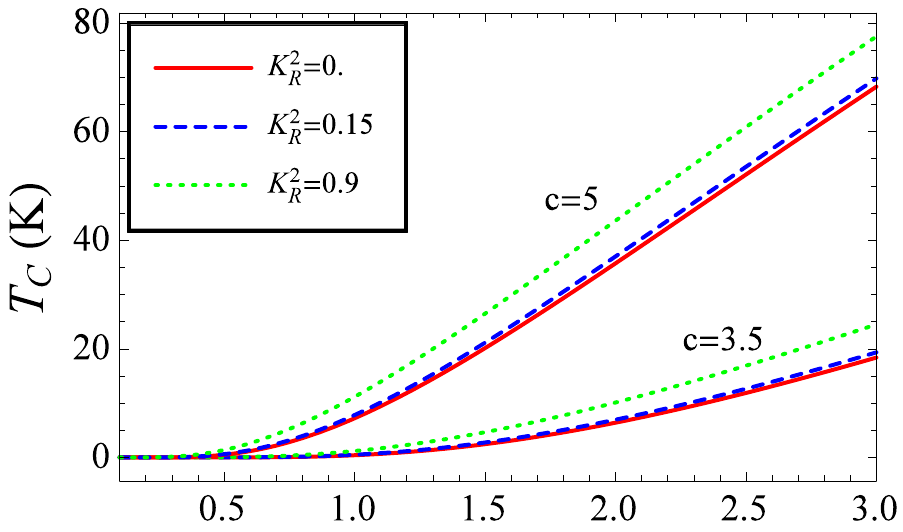}
\includegraphics[width=0.9\linewidth]{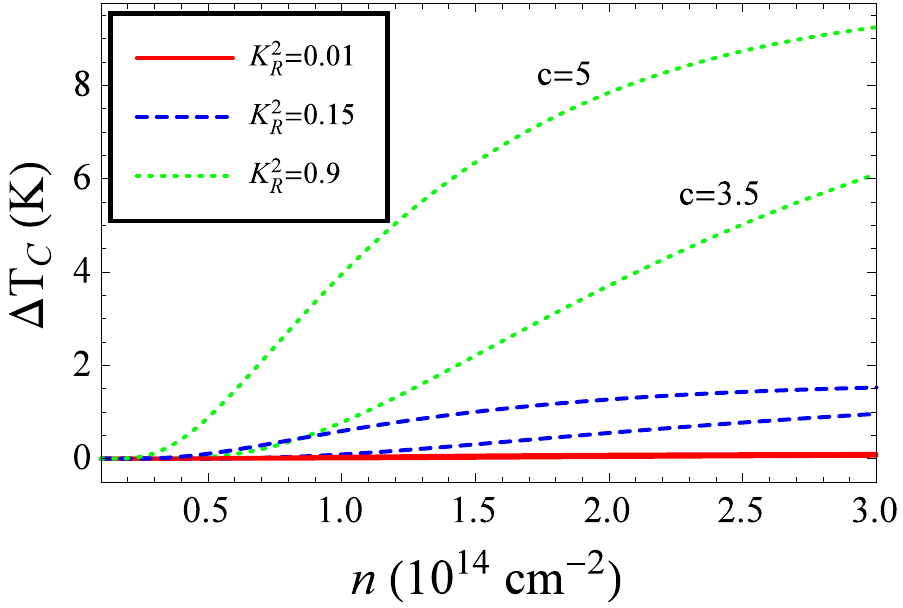}
\caption{(Color online) Critical temperature and variation from the \textquotedblleft bare\textquotedblright~one in \cite{Einenkel2011} for graphene on a piezoelectric substrate, as a function of the conduction band density. Three pairs of plots are given for three different values of $K_R$ and the two values of the constant $C = 3.5$ and 5 in Eq. (5.1) for $\lambda$ in the previous reference \cite{Einenkel2011}. }
\label{fig:tcPlot}
\end{figure}

The upshot of this discussion is that the long-wavelength piezoelectric phonons work in favor of pairing instabilities, as shown in Fig. \ref{fig:tcPlot}. We emphasize, however, that we are not claiming that a piezoelectric substrate {\it per se} necessarily increases the critical temperature, since it could be the case that other piezoelectric fluctuations not considered in the present study (e.g. shorter wavelength modes) could work against pairing instabilities.

\section{Conclusions}
In conclusion, we have derived a general expression for the two-dimensional electron-phonon piezoelectric interaction valid for any piezoelectric substrate covered by a two-dimensional electron system, as in the classical 2D Fr\"ohlich hamiltonian for the optical phonons, and characterized the magnitude of the interaction. Our results show that electron overscreening cannot be achieved just with the strongest piezoelectric phonons within our assumptions because $K_R^2<1$ is always satisfied. Nevertheless, these phonons could help further in other contexts where the 2D superconductivity is known to exist, for example in bulk few-layer $\text{MoS}_{2}$ with most of the carriers confined to the first layer \cite{Ye2012,Roldan2013}; or postulated to exist but not yet observed due to experimental difficulties (e.g. very heavily doped graphene \cite{Einenkel2011}). Other example is the recent high-temperature superconductor system of 2D FeSe on top of the ferroelectric $\text{SrTiO}_{3}$, whose optical phonons have been analyzed arriving at conclusions similar to ours \cite{Gorkov2015}, and where the strong piezoelectric phonons could play a role as well.
\begin{acknowledgments}
We want to thank Fernando Calle and J\"urgen Schiefele for valuable discussions. DGG acknowledges financial support from Campus de Excelencia Internacional (Campus Moncloa UCM-UPM). This work has been supported by the Spanish Ministry of Economy (MINECO) through Grants No. FIS2011-23713, FIS2013-41716-P; the European Research Council Advanced Grant (contract 290846), and the European Commission under the Graphene Flagship, contract CNECTICT- 604391.

\end{acknowledgments}

\appendix

\section{Piezo-SAW phonon-electron interaction vertex}
\label{app:PiezoElectricVertex}

The situation is depicted in Fig.~\ref{fig:PiezoSAWSchema}. The $z=0$ interfacial surface is supposed to be free of tension and, when acting as a substrate to a deposited 2D electronic material sheet, free of electrodes as well. However, in the present section, we will allow for flat electrodes (at $z=0^{+}$) which supply no mechanical stresses.
The purpose of this section is to show that the interaction between the propagating piezoelectric SAWs and the electrons of the 2D electronic sheet can be described with a Hamiltonian of the form Eq.~(\ref{eqn:PAHamiltonian}) with (we use SI units throughout the present section, as it is typical for piezoelectrics):
\begin{align}
\omega_\mathbf{q} 		&= v_s(\theta) |\mathbf{q}| ~, \nonumber \\
\gamma^{\text{PA}}_\mathbf{q} 	&= \frac{K_R(\theta)}{\sqrt{2}} 
\left[
	\frac{v_s(\theta)\hbar e^2}{(\tilde{\varepsilon}_{\text{HF}}(\theta)+1)\varepsilon_\text{vac}}
		\right]^{1/2} ~,
\label{eqn:PAInteractionVertex}
\end{align}
where we have written $\mathbf{q}:=q\left[\cos(\theta),\sin(\theta)\right]$ and $\varepsilon_{\text{vac}}$ is the air or vacuum electric permittivity. The piezoelectric specific parameters are $v_s(\theta)$, the piezoelectric SAW velocity; $0\le K_R(\theta)^2<1$, the {\it SAW electromechanical coupling coefficient}; and $\tilde{\varepsilon}_{\text{HF}}(\theta):=\tilde{\varepsilon}(\mathbf{q}/\omega), \omega\rightarrow\infty$ (in the acoustic frequency scale), the high-frequency (HF) limit of the {\it piezoelectric surface permittivity} (see \cite{Royer2000,Royer2000a}). They all depend on the propagation direction of the SAW, as the notation suggests.

\subsection{Piezoelectric Surface Acoustic Waves}

For a general introduction to piezoelectric SAWs see Refs. \cite{Royer2000,Royer2000a,Farnell1970,Morgan2010}. The point displacement, $u_i(\mathbf{r},t)$, where $i=1,2,3$ for the $x,y,z$ directions respectively in the piezoelectric substrate, obeys the elastic equation of motion (in the present appendix, it is used implicit sums on repeated indexes):
\begin{equation}
\frac{\partial^2 u_i}{\partial t^2}=\frac{\partial T_{ij}}{\partial x_j},
\label{eqn:elasticNewton}
\end{equation}
where $T_{ij}(\mathbf{r},t)$ is the symmetrical stress tensor. Poisson's equation for the electric displacement $D_i(\mathbf{r},t) $ (no charges inside the material) is:
\begin{equation}
\frac{\partial D_i}{\partial x_i}=0 ~.
\label{eqn:Poisson}
\end{equation}

The coupled constitutive (linear) equations relate the stress tensor and electric displacement with the strain tensor and electric field (here written as the gradient of the electric potential $E_i=-\partial \varphi/\partial x_i$)
\begin{align}
T_{ij} 	&= c_{ijkl}\partial u_k/\partial x_l+e_{kij} \partial \varphi/\partial x_k \nonumber \\
D_i		&= -\varepsilon_{jk}\varepsilon_\text{vac}\partial \varphi/\partial x_j+e_{ijk} \partial u_j/\partial x_k ~,
\label{eqn:piezoConstitutive}
\end{align}
where we have introduced the elastic constant tensor $c_{ijkl}\equiv c_{ijkl}^E$ measured at constant electric field, the electric (relative) permittivity tensor $\varepsilon_{ij}\equiv \varepsilon_{ij}^S$ measured at constant strain and
the piezoelectric tensor $e_{ijk}$.

The SAWs are solutions to (\ref{eqn:elasticNewton}-\ref{eqn:piezoConstitutive}) in the form of plane waves propagating along the surface $z=0$ in the direction specified by $\mathbf{b}=\left[\cos\theta,\sin\theta\right]$
\begin{align}
u_j		&= \alpha_j \exp [ik(b_ix_i-vt)] \nonumber \\
\varphi &= \alpha_4 \exp [ik(b_ix_i-vt)] ~,
\end{align}
and we have extended here to 3D the definition of $\textbf{b}:=\left[\cos(\theta),\sin(\theta),b_3\right]$ so that $b_3$ is now a variable to be determined by the requirements of boundedness or causality of normal modes (see below). In what follows, we assume always $v>0$ and $k>0$.

The resulting linear equations for the amplitudes $\alpha_a$, (here $a,b=1,2,3,4$ and $i,j,k,l=1,2,3$) are:
\begin{align}
	0 &= \left( \Gamma_{ab}-\delta'_{ab}\rho v^2 \right)\alpha_a \nonumber \\
	\Gamma_{jk} &= b_ib_lc_{ijkl} \nonumber \\
	\Gamma_{j4} &= b_ib_ke_{ijk} \nonumber \\
	\Gamma_{44} &= -b_ib_k\varepsilon_{ik}\varepsilon_\text{vac} ~,
\label{eqn:normalModesAmplitudes}
\end{align}
with $\delta'_{ij}=\delta_{ij}, ~\delta'_{4a}=\delta'_{a4}=0$, and $\rho$ the constant density of the piezoelectric solid.

Note that $k$ disappears, which means that there is no dispersion for a given propagating direction. Hence, given the propagation direction $\theta$ and the velocity $v$, the solutions for $\det\left( \Gamma_{ab}-\delta'_{ab}\rho v^2 \right)=0$ as a function of $b_3$ is a set of no more than 8 complex values, in which, because of the reality of the coefficients, each complex root comes together with its conjugate, and among these we have to choose the ones with $\text{Im}\,b_3<0$, so that the modes are not exponentially growing deep into the solid. In the case of purely real solutions, usual arguments on causality demand that we have to take only those modes with radiation (outgoing from the surface $z=0$) boundary conditions $db_3(v)/dv<0$ (see \cite{Hashimoto2000}). Hence, the total number of allowed modes is 4, and the general solution we write as (we use now $u_4:=\varphi$ and write somehow loosely $\mathbf{r}=(\mathbf{R},z)$, with the 2D $\mathbf{R}=(x,y)$):
\begin{equation}
	u_a(\mathbf{r},t) = C_n \alpha_a^{(n)} e^{ikb_3^{(n)}z} \exp\left[ik(\textbf{b}\cdot \textbf{R}-vt)\right] ~,
\label{eqn:PSAWWaveMaterial}
\end{equation}
with $n=1...4$ indexing the normal modes.

Much simpler is the equation at vacuum/air. The solution is purely electric and can be written as:
\begin{equation}
	\varphi(\mathbf{R},z,t)=u_4(\mathbf{R},0,t) e^{-kz} ~,
\label{eqn:PSAWWaveVacuum}
\end{equation}
because of continuity of the potential.

The mechanical boundary condition at the interface $T_{i3}(\mathbf{R},0,t)=0$ leads to (here
$\textbf{b}^{(n)}:=\left[\cos(\theta),\sin(\theta),b_3^{(n)}\right]$):
\begin{equation}
	C_n b^{(n)}_k (\alpha_j^{(n)} c_{3ijk}+\alpha_4^{(n)}e_{k3i})=0 ~,
\label{eqn:mechanicalBoundaryCondition}
\end{equation}
hence $C_i$ are proportional to $C_4$.

The normal component of the electric displacement is, at the interface:
\begin{align}
	D_3(\mathbf{R},0^-,t) =& ik\exp\left[ik(\textbf{b}\cdot\textbf{R}-vt)\right] \nonumber \\
	& \times C_n b^{(n)}_k (\alpha_j^{(n)} e_{3jk}-\alpha_4^{(n)}\varepsilon_{3k}\varepsilon_\text{vac}) ~,
\end{align}
and this allows to introduce the {\it piezoelectric surface permittivity} as the ratio:
\begin{align}
	\tilde{\varepsilon}(\textbf{k}/\omega) &:= -\frac{D_3(\mathbf{R},0^-,t)}{k \varphi(\mathbf{R},0^-,t)\,\varepsilon_\text{vac}} \nonumber \\
	&= -i \frac{C_n b^{(n)}_k (\alpha_j^{(n)} e_{3jk}-\alpha_4^{(n)}\varepsilon_{3k})}{C_m \alpha_4^{(m)}\,\varepsilon_\text{vac}} ~,
\label{eqn:piezoelectricSurfacePermittivity}	
\end{align}
which only depends on $v$ and $\theta$, through the relations $\textbf{k}:=k \textbf{b}$ and $\omega:=kv$.

Similarly, on the other side of the interface we have the obvious relation
\begin{equation}
	1 = \frac{D_3(\mathbf{R},0^+,t)}{k \varphi(\mathbf{R},0^+,t)\,\varepsilon_\text{vac}} ~.
\end{equation}
Hence, the surface charge at the interface  can be expressed as:
\begin{equation}
	\sigma(0)=D_3(0^+)-D_3(0^-)=k \varphi(0) [1+\tilde{\varepsilon}(\textbf{k}/\omega)]\varepsilon_\text{vac} ~,
\label{eqn:surfaceChargePSAW}
\end{equation}
where the dependence $\exp\left[ik(\textbf{b} \cdot \textbf{r}-vt)\right]$ is implicitly assumed and the electrodes should be placed perpendicular to the propagation direction.

From Eq.~(\ref{eqn:surfaceChargePSAW}), a source free propagating wave only exists if
\begin{equation}
1+\tilde{\varepsilon}(\textbf{k}/\omega)=0~,
\label{eqn:RayleighWaveCondition}
\end{equation}
i.e. the phase velocity $v_s(\theta)$ of the wave is given by $1+\tilde{\varepsilon}(\textbf{b}/v_s(\theta))=0$. This is the piezoelectric Rayleigh waves condition.

In \cite{Lothe1976} it is shown that the energetic stability of the piezoelectric guarantees that $\text{Im}\,\tilde{\varepsilon}(\textbf{b}/v_s(\theta))=0$ up to a $v_L(\theta)>v_0(\theta)$, with $\tilde{\varepsilon}(\textbf{b}/v_0(\theta))=0$. In that range, the four modes in Eq.~(\ref{eqn:PSAWWaveMaterial}) are purely decaying on the substrate side. $v_L(\theta)$ marks the starting point at which the {\it piezoelectric surface permittivity} has an imaginary part, which reflects the influence of bulk modes.

\subsection{Hamiltonian and interaction vertex}\label{app:HamiltonianInterVertex}

The linear equations of piezoelectricity, Eqs.~(\ref{eqn:elasticNewton}-\ref{eqn:piezoConstitutive}), can be derived from a Lagrangian (see \cite{Tiersten1969})
\begin{align}
	L\left[u_j, \varphi\right] &= \frac{1}{2} \int d^3\mathbf{r}\left[\rho \dot{u}_i\dot{u}_i -c_{ijkl}u_{i,j}u_{k,l} \right. \nonumber \\
	&- \left. 2e_{ijk}\varphi_{,i}u_{j,k} +\varepsilon_{ij}\varepsilon_\text{vac}\varphi_{,i}\varphi_{,i}\right] ~,
\end{align}
where we have written $_{,j}:=\partial/\partial x_j$ and $\dot{u}_i:=\partial u_i/\partial t$. The canonical momentum to $\varphi$ is zero, so that the system is constrained. The Hamiltonian is then
\begin{align}
	H\left[u_j, \varphi\right] &=  \frac{1}{2} \int d^3\mathbf{r}\left(\rho \dot{u}_i\dot{u}_i +c_{ijkl}u_{i,j}u_{k,l} +\varepsilon_{ij}\varepsilon_\text{vac}\varphi_{,i}\varphi_{,i}\right).
\label{eqn:PSAWHamiltonian}
\end{align}

For a given harmonic propagating (no surface charges) piezoelectric SAW, i.e., a wave with the form of $\text{Re}\,u_a(\textbf{r},z,t)$ from Eqs.~(\ref{eqn:PSAWWaveMaterial}-\ref{eqn:PSAWWaveVacuum}) fulfilling the equations of motion Eqs.~(\ref{eqn:elasticNewton}-\ref{eqn:piezoConstitutive}) and boundary conditions Eqs.~(\ref{eqn:mechanicalBoundaryCondition}-\ref{eqn:surfaceChargePSAW}) with $\sigma(0)=0$, it is straightforward to show that the kinetic energy (first term in Eq.~(\ref{eqn:PSAWHamiltonian}), coming exclusively from elastic vibrations in the substrate) is the same as the potential energy (last two terms in Eq.~(\ref{eqn:PSAWHamiltonian}), contains contributions from elastic deformation and electrostatic stored energy both in the substrate and in free space). On the other hand \cite{Royer2000}, for the interval $0<v_s(\theta)<v_L(\theta)$, positivity of the kinetic and potential energies give $\partial \tilde{\varepsilon}(\textbf{k}/\omega)/\partial \omega>0$. For these kind of waves we have that \cite{Royer2000} (when $1+\tilde{\varepsilon}(\textbf{k}/\omega)=0$)
\begin{align}
	H_{\text{harm}} &=  \frac{1}{4} A k \omega \frac{\partial \tilde{\varepsilon}(\textbf{k}/\omega)\varepsilon_\text{vac}}{\partial \omega} |\varphi_0|^2
	\nonumber \\
	&= \frac{1}{2} Ak|\varphi_0|^2 \frac{(\tilde{\varepsilon}_{\text{HF}}(\theta)+1)\varepsilon_\text{vac}}{K_R^2(\theta)} ~,
\label{eqn:HamiltonianHarmonicPSAW}
\end{align}
where $A$ is the area of the sample, $\varphi_0:=C_n\alpha_4^{(n)}$ is the amplitude of the electric potential at the interface (see Eq.~(\ref{eqn:PSAWWaveMaterial})), and we have introduced the high-frequency limit $\tilde{\varepsilon}_{\text{HF}}(\theta):=\tilde{\varepsilon}(\mathbf{k}/\omega), \omega\rightarrow\infty$ and the \textit{SAW electromechanical coupling coefficient}, $K_R(\theta)$ through the relation at $1+\tilde{\varepsilon}(\textbf{k}/\omega)=0$:
\begin{equation}
	\frac{K_R^2(\theta)/2}{\tilde{\varepsilon}_{\text{HF}}(\theta)+1} = \left[\omega\frac{\partial \tilde{\varepsilon}(\textbf{k}/\omega)}{\partial \omega}\right]^{-1} ~.
\end{equation}

The electrons of the graphene sheet (or any other charged two dimensional structure deposited at the piezoelectric substrate) feel the electric potential of the piezoelectric SAW. The interaction is then the total potential at the position of the electron
\begin{equation}
	V_{\text{PA}}(\textbf{R})=-e\, \varphi(\textbf{R},0,t=0) ~.
\end{equation}

On the other hand, the one-phonon normalization means that $\varphi_0$ from Eq.~(\ref{eqn:HamiltonianHarmonicPSAW}) should be chosen so that $H_{\text{harm}}=\hbar \omega= \hbar v_s(\theta) k$, and thus we finally get the hamiltonian Eq. (\ref{eqn:PAVertex}).

\subsection{Response functions}

We now consider a 1D situation, in which flat electrodes parallel to the $y$-axis operate on top of the piezoelectric substrate shown in Fig.~\ref{fig:PiezoSAWSchema}. Therefore, we chose $\theta=0$ and there is no $y$ dependence. We omit to write $\theta$ in this subsection.

The charge-potential relation (\ref{eqn:surfaceChargePSAW}) for the amplitudes is written so that we define the complex admittance $\chi(k,\omega)$ as:
\begin{align}
	\varphi(k,\omega) &= \gamma(k,\omega) \sigma(k,\omega) \nonumber \\
	\gamma(k,\omega) &:= \frac{1}{|k|} \frac{1}{(\tilde{\varepsilon}(k,\omega)+1)\varepsilon_\text{vac}} ~,
\label{eqn:admittanceDefinition}
\end{align}
where we have to allow now for the possibility of negative $k$, because we are omitting the $\theta$ dependence. From Eq.~(\ref{eqn:normalModesAmplitudes}), $\tilde{\varepsilon}(k,\omega)=f((\omega/k)^2)=f(v^2)$, and its analytical extensions can be guessed from the requirements of causality, which for $\omega>0$ means that the poles and zeros of $\gamma(k,\omega)$ are placed in the lower complex $\omega$ half-plane.

We define the instantaneous part
\begin{align}
	\gamma_\infty(k) &:= \frac{1}{|k|} \frac{1}{(\tilde{\varepsilon}_\text{HF}+1)\varepsilon_\text{vac}}
	\nonumber \\
	 &= \int dx\,e^{-ikx} \gamma_\infty(x) ~,
\label{eqn:admittanceInstantaneous}
\end{align}
and the retarded and static contributions
\begin{align}
	\gamma_{\text{ret}}(k,\omega) &:= \gamma(k,\omega)-\gamma_\infty(k)
	\nonumber \\
	&= \int dx \int_0^{\infty} e^{i(\omega t - kx)} \phi(x,t) ~,
\label{eqn:admittanceRetarded} \\
	\gamma_0(k) &:= \gamma(k,0)=\frac{1}{|k|} \frac{1}{(\tilde{\varepsilon}_{\text{LF}}+1)\varepsilon_\text{vac}}
	\nonumber \\
	&= \int dx e^{-ikx} \gamma_0(x)
	\nonumber \\
	\gamma_0(x) &= \gamma_\infty(x)+\int_0^\infty ds \phi(x,s) e^{-\eta s} ~,
\label{eqn:admittanceStatic}
\end{align}
where $\tilde{\varepsilon}_{\text{LF}} :=\tilde{\varepsilon}(\mathbf{k}/\omega), \omega\rightarrow 0$ and 
$\eta$ is to be understood as $\eta \rightarrow 0^+$.

All this amounts to writing the general linear causal relation \cite{Kubo2012}
\begin{align}
	\varphi(x,t) &= \int dx' \left[\gamma_\infty(x-x')\sigma(x',t) \right.
	\nonumber \\
	 &+ \left.\int_{-\infty}^{t} dt'\,\phi(x-x',t-t') \sigma(x',t')\right] ~.
\label{eqn:linearResponse}
\end{align}

The power delivered to the electrodes to maintain a given $\varphi(x,t),\sigma(x,t)$ (in this subsection we assume that all fields which depend on space-time are real) is:
\begin{equation}
	\frac{dU(t)}{dt}=\sqrt{A} \int dx\,\varphi(x,t) \dot{\sigma}(x,t) ~,
\label{eqn:powerDelivered}
\end{equation}
where $\sqrt{A}$ is the length along the $y$-direction.

If starting from zero fields and charges, we adiabatically turn on a given surface charge distribution $\sigma(x,t)=\sigma(x) \exp(\eta t)$, from Eqs.(\ref{eqn:linearResponse}-\ref{eqn:powerDelivered}), the total energy supplied is:
\begin{align}
	\frac{\Delta U_{\text{ad}}}{\sqrt{A}} &= \int dx \int dx' \sigma(x)\frac{\gamma_0(x-x')}{2} \sigma(x')
	\nonumber \\
	& = \frac{1}{2(\tilde{\varepsilon}_{\text{LF}}+1)\varepsilon_\text{vac}}\int \frac{dk}{2\pi} \frac{|\sigma(k)|^2}{|k|} ~.
\label{eqn:adiabaticEnergy}
\end{align}

Analogously, an instantaneous charging to the same final charge distribution $\sigma(x,t)=\theta_\epsilon(t)\sigma(x)$, with $\theta_\epsilon(t)$ a differentiable approximation to the Heaviside $\theta$-function such that $\theta_\tau(t)\rightarrow\theta(t), \tau\rightarrow 0^+$, requires an amount of work given by:
\begin{align}
	\frac{\Delta U_{\text{inst}}}{\sqrt{A}} &= \int dx \int dx'\,\sigma(x)\frac{\gamma_\infty(x-x')}{2} \sigma(x')
	\nonumber \\
	& = \frac{1}{2(\tilde{\varepsilon}_{\text{HF}}+1)\varepsilon_\text{vac}}\int \frac{dk}{2\pi}  \frac{|\sigma(k)|^2}{|k|} ~.
\label{eqn:instantaneousEnergy}
\end{align}

The second process being non-adiabatic, it absorbs more energy from the source that exerts a work on the system. This extra energy is employed in inducing surface and bulk wave excitations. As a result, $\Delta U_{\text{inst}}>\Delta U_{\text{ad}}$, which implies
\begin{equation}
\tilde{\varepsilon}_{\text{HF}}<\tilde{\varepsilon}_{\text{LF}} ~.
\label{eqn:eps-inequality}
\end{equation}

After the sudden charge, i.e. at $t>0$, the time evolution and relaxation of the potential are, due to Eqs.~(\ref{eqn:admittanceStatic},\ref{eqn:linearResponse}):
\begin{align}
	\varphi(x,t) &= \int dx' \sigma(x')\left[\gamma_\infty(x-x')+ \int_{0}^{t} dt' \phi(x-x',t-t') \right]
	\nonumber \\
	& \xrightarrow{t\rightarrow\infty}
	\int dx' \sigma(x')\gamma_0(x-x') ~,
\label{eqn:timeEvolution_RelaxationPotential}
\end{align}
this relaxed field being the same as that obtained after the adiabatic process to the same charge distribution.

The space Fourier, time Fourier-Laplace transform of this potential is:
\begin{align}
	 \varphi(k,\omega) &:= \int dx \int_0^\infty dt\, e^{i(\omega t - kx)} \varphi(x,t)
	 \nonumber \\
	 &= \frac{i\sigma(k)}{\omega+i\eta}\gamma(k,\omega) ~,
\end{align}
where the change $\omega \rightarrow \omega + i\eta$ $(\eta \equiv 0^+)$ is made to ensure convergence.

As $\gamma(k,\omega)$ has poles at the Rayleigh waves condition (\ref{eqn:RayleighWaveCondition}), we can isolate their contribution, $\varphi_{\text{RW}}(x,t)$ to $\varphi(x,t)$,
\begin{align}
	\varphi_{\text{RW}}(k,\omega) &:=\frac{i\sigma(k)}{|k|} \frac{K_R^2/2}{(\tilde{\varepsilon}_{\text{HF}}+1)\varepsilon_\text{vac}}
	\nonumber \\
	& \left(\frac{1}{\omega-\omega_k+i\eta}+\frac{1}{\omega+\omega_k+i\eta}\right) ~,
\end{align}
where $\omega_k=vk$ and the two terms come from the two identical SAWs propagating to the right and left. A small $0^+$ has been added to ensure that the poles of the admittance are in the lower complex $\omega$ half-plane. Inverting to get the spacetime behavior, we obtain two dispersionless propagating SAWs:
\begin{equation}
	\varphi_{\text{RW}}(x,t)=\frac{K_R^2}{2}\left[\varphi(x-vt,0^+)+\varphi(x+vt,0^+)\right] ~,
\end{equation}
where $\varphi(x\pm vt,0^+)=\int (dk/2\pi)\,e^{ik(x\pm vt)} \gamma_\infty(k) \sigma(k)$ [see Eq.~(\ref{eqn:timeEvolution_RelaxationPotential})]. The energy carried by these two pulses is, using Eq.~(\ref{eqn:energyCarried_PotentialPulse}):
\begin{equation}
	\Delta U_{\text{RW}}=\sqrt{A}\frac{K_R^2/2}{(\tilde{\varepsilon}_{\text{HF}}+1)\varepsilon_\text{vac}}\int \frac{dk}{2\pi} \frac{|\sigma(k)|^2}{|k|} ~,
\end{equation}
which is the energy stored in each traveling SAW, i.e. $\Delta U_\text{RW}=K_R^2\,\Delta U_\text{inst}$ from Eq. (\ref{eqn:instantaneousEnergy}). Since we have at our disposal no more than $\Delta U_{\text{inst}}-\Delta U_{\text{ad}}>0$, the condition $\Delta U_{\text{RW}}<\Delta U_{\text{inst}}-\Delta U_{\text{ad}}$ must be fulfilled. From Eqs.(\ref{eqn:adiabaticEnergy}-\ref{eqn:eps-inequality}) we conclude that:
\begin{equation}
	K_R^2 \le \frac{\tilde{\varepsilon}_{\text{LF}}-\tilde{\varepsilon}_{\text{HF}}}{\tilde{\varepsilon}_{\text{LF}}+1} < 1 ~. \label{eqn:kSquareBound}
\end{equation}

\subsection{High frequency limit of $\tilde{\varepsilon}(\mathbf{k}/\omega)$} \label{app:HFLFPSAWPermittivity}

In this section we want to show that, if we take the propagating direction along $x$-axis, then:
\begin{equation}
\tilde{\varepsilon}_\text{HF}=\varepsilon_{p}:= \sqrt{\varepsilon_{11}\varepsilon_{33}-(\varepsilon_{13})^2} ~.
\end{equation}

In fact, we write the modes equation Eq. (\ref{eqn:normalModesAmplitudes}) as,
\begin{equation}
	\hat{M}
	\left(
	\begin{array}{c}
	\vec{u}\\
	\varphi\\
	\end{array}
	\right)
	\equiv
	\left(
	\begin{array}{cc}
	\Gamma-\rho v^2 \textbf{1} & \vec{\gamma} \\
	\vec{\gamma}^\top & -\varepsilon\varepsilon_\text{vac} \\
	\end{array}
	\right)
	\left(
	\begin{array}{c}
	\vec{u}\\
	\varphi\\
	\end{array}
	\right)
	=0	~,
\label{eqn:normalModesMatrixForm}
\end{equation}
where the form of the $3\times3$ matrix $\Gamma$, $3\times1$ vector $\vec{\gamma}$ and constant $\varepsilon$ as a function of $b$ (where $\textbf{b}=(1,0,b)$) can be read from Eq. (\ref{eqn:normalModesAmplitudes}).

There are two possibilities for the variation of $b$ as $v\rightarrow\infty$, either (a) $b \rightarrow \ b_{\text{sm}}<\infty$, (\textquotedblleft sm\textquotedblright~ means small) or (b) $b \sim  b_{\text{bg}} \rightarrow \infty$ (\textquotedblleft bg\textquotedblright~ is for big).

In case (a), $\Gamma-\rho v^2 \textbf{1}$ will never be singular, so using the determinant formula from Schur's complement $\det(\hat{M})=\det(\Gamma-\rho v^2 \textbf{1})\det(-\varepsilon -\vec{\gamma}\cdot(\Gamma-\rho v^2 \textbf{1})^{-1}\cdot\vec{\gamma})$, it is immediate to realize that $\varepsilon =0+O(v^{-2})$, which leads to the decaying root $b_{\text{sm}}=-(\varepsilon_{31}+i \varepsilon_{p})/\varepsilon_{33}$.

From the modes equation (\ref{eqn:normalModesMatrixForm}), we find that:
\begin{equation}
	\left(
	\begin{array}{c}
	\vec{u}_{\text{sm}}\\
	\varphi_{\text{sm}}\\
	\end{array}
	\right)
	\simeq
	\left(
	\begin{array}{c}
	O(v^{-2})\\
	1\\
	\end{array}
	\right) ~,
\end{equation}
where here and in the rest of this subsection, we normalize the modes amplitudes so that $\varphi_{\text{sm,bg}}=1$.

For the other case (b), from the modes equation (\ref{eqn:normalModesMatrixForm}) we find that $b_{\text{bg}}=O(v)$, hence, expanding $\hat{M}$ from:
\begin{align}
	\Gamma_{ij} &\simeq b_{\text{bg}}^2 c_{3ij3}
	\nonumber \\
	\gamma_i &\simeq e_{33i}  b_{\text{bg}}^2
	\nonumber \\
	\varepsilon &\simeq \varepsilon_{33} b_{\text{bg}}^2 ~,
\label{eqn:asymptoticCoefficientsMMatrix}
\end{align}
but now the general form of these modes is
\begin{equation}
\left(
\begin{array}{c}
\vec{u}_{\text{bg}}\\
\varphi_{\text{bg}}\\
\end{array}
\right)
\simeq
\left(
\begin{array}{c}
\alpha^{(i)}_j\\
1\\
\end{array}
\right) ~,
\end{equation}
where we have used the notation in Eq. (\ref{eqn:PSAWWaveMaterial}) and chosen $\alpha^{(1,2,3)}_a$ for the three $(\vec{u}_{\text{bg}},\varphi_{\text{bg}})$ modes and $\alpha^{(4)}_a$ for the $(\vec{u}_{\text{sm}},\varphi_{\text{sm}})$ mode.

Choosing the constant $C_4=1$, the mechanical boundary condition (\ref{eqn:mechanicalBoundaryCondition}) leads to:
\begin{equation}
	0\simeq C_kb^{(k)}(\alpha^{(k)}_jc_{3ij3}+e_{33i})+(e_{13i}+b^{(4)}e_{33i}) ~,
\end{equation}
and $C_k=O(v^{-1})$, so the denominator in Eq.~(\ref{eqn:piezoelectricSurfacePermittivity}) can be approximated as $C_m \alpha_4^{(m)}\simeq 1$.

On the other hand, the %$|_{\text{bg}}$
``big'' (bg) contribution to the displacement field is, to order $O(v^0)$:
\begin{equation}
	D_3(0^-)|_{\text{bg}} \simeq ik c_ib^{(i)}(\alpha^{(i)}_j e_{33j}-\varepsilon_{33}\varepsilon_\text{vac})\simeq 0 ~,
\end{equation}
the last approximate equality comes from the second Eq.~(\ref{eqn:normalModesMatrixForm}) together with Eq.~(\ref{eqn:asymptoticCoefficientsMMatrix}).

Collecting all these results together with the ``small'' (sm) contribution to $D_3(0^+)$ into Eq.~(\ref{eqn:piezoelectricSurfacePermittivity}), we finally get \cite{Darinskii2007}:
\begin{equation}
	\tilde{\varepsilon}_{\text{HF}}=-i b^{4}_k(\varepsilon_{3k})=\varepsilon_{p} ~.
\end{equation}

\subsection{Energy carried by the piezoelectric SAW pulse}
\label{app:PSAWpulseEnergy}

For piezoelectric phenomena, the Poynting vector is (see \cite{Royer2000}):
\begin{equation}
	P_j=-T_{ij} \dot{u_i}+\varphi \dot{D_j} ~,
\end{equation}
which, after use of Eq. (\ref{eqn:piezoConstitutive}) can be seen to be a bilinear expression in the vectors $(u_a,u_{a,i})$ and $(\dot{u}_b,\dot{u}_{b,j})$ (here $i,j=1,2,3$ and $a,b=1,2,3,4$; where $u_4=\varphi$). For a given pulse propagating in the $x$-direction, $u_a(x,y,z,t)=f_a(x-vt,z)$, we are interested in the total energy which crosses $x$ (is obviously independent of $x$)
\begin{align}
	\Delta U_{\text{pulse}}	&= \iiint\limits_{-\infty}\limits^{\infty} dt\,dy\,dz \,P_1(x,y,z,t)
	\nonumber \\
	&= \sqrt{A}\iint\limits_{-\infty}\limits^{\infty} dt \,dz \, g_r(x-vt,z) P_{rs} g_s(x-vt,z) ~,
\end{align}
where $g_{r/s}$ are taken from the components $u_a,u_{a,i},\dot{u}_a,\dot{u}_{a,i}$ with $r,s=1,2,...,16$, and $P_{rs}$ is a constant matrix with elements of the tensors $\hat{e},\hat{c},\hat{\varepsilon}$. Fourier analyzing $g_r(x-vt,z)=\int (dk/2\pi)\,e^{ik(x-vt)}g_r(k,z)$, where because of reality $g_r(k,z)^*=g_r(-k,z)$, we obtain:
\begin{equation}
	\Delta U_{\text{pulse}} =\sqrt{A}\frac{1}{v} \int \frac{dk}{2\pi}\,\int dz\,g_r(k,z) P_{rs} g_s(k,z)^* ~,
\end{equation}
but then \cite{Royer2000,Morgan2010}:
\begin{equation}
	\frac{1}{2}\int dz\,g_r(k,z) P_{rs} g_s(k,z)^*=\frac{v|k|}{4}\omega \frac{\partial \tilde{\varepsilon}(\textbf{k},\omega)}{\partial \omega} |\varphi(k,0)|^2
\end{equation}
is the time-average power per unit length crossing a $yz$-section by a harmonic piezoelectric SAW, whose electric potential amplitude is $\varphi(k,0)$ at the interface. The result is:
\begin{equation}
	\Delta U_{\text{pulse}}= \sqrt{A}\,\frac{(\tilde{\varepsilon}_{\text{HF}}+1)\varepsilon_\text{vac}}{K_R^2} \int \frac{dk}{2\pi} |k| |\varphi(k,0)|^2 ~.
\label{eqn:energyCarried_PotentialPulse}
\end{equation}

\bibliographystyle{apsrev4-1}
\bibliography{PiezoGraphene}

\begin{thebibliography}{38}%
\makeatletter
\providecommand \@ifxundefined [1]{%
 \@ifx{#1\undefined}
}%
\providecommand \@ifnum [1]{%
 \ifnum #1\expandafter \@firstoftwo
 \else \expandafter \@secondoftwo
 \fi
}%
\providecommand \@ifx [1]{%
 \ifx #1\expandafter \@firstoftwo
 \else \expandafter \@secondoftwo
 \fi
}%
\providecommand \natexlab [1]{#1}%
\providecommand \enquote  [1]{``#1''}%
\providecommand \bibnamefont  [1]{#1}%
\providecommand \bibfnamefont [1]{#1}%
\providecommand \citenamefont [1]{#1}%
\providecommand \href@noop [0]{\@secondoftwo}%
\providecommand \href [0]{\begingroup \@sanitize@url \@href}%
\providecommand \@href[1]{\@@startlink{#1}\@@href}%
\providecommand \@@href[1]{\endgroup#1\@@endlink}%
\providecommand \@sanitize@url [0]{\catcode `\\12\catcode `\$12\catcode
  `\&12\catcode `\#12\catcode `\^12\catcode `\_12\catcode `\%12\relax}%
\providecommand \@@startlink[1]{}%
\providecommand \@@endlink[0]{}%
\providecommand \url  [0]{\begingroup\@sanitize@url \@url }%
\providecommand \@url [1]{\endgroup\@href {#1}{\urlprefix }}%
\providecommand \urlprefix  [0]{URL }%
\providecommand \Eprint [0]{\href }%
\providecommand \doibase [0]{http://dx.doi.org/}%
\providecommand \selectlanguage [0]{\@gobble}%
\providecommand \bibinfo  [0]{\@secondoftwo}%
\providecommand \bibfield  [0]{\@secondoftwo}%
\providecommand \translation [1]{[#1]}%
\providecommand \BibitemOpen [0]{}%
\providecommand \bibitemStop [0]{}%
\providecommand \bibitemNoStop [0]{.\EOS\space}%
\providecommand \EOS [0]{\spacefactor3000\relax}%
\providecommand \BibitemShut  [1]{\csname bibitem#1\endcsname}%
\let\auto@bib@innerbib\@empty
%</preamble>
\bibitem [{\citenamefont {Landau}\ and\ \citenamefont
  {Lifshitz}(1986)}]{Landau1986}%
  \BibitemOpen
  \bibfield  {author} {\bibinfo {author} {\bibfnamefont {L.~D.}\ \bibnamefont
  {Landau}}\ and\ \bibinfo {author} {\bibfnamefont {E.~M.}\ \bibnamefont
  {Lifshitz}},\ }\href@noop {} {\emph {\bibinfo {title} {Theory of
  elasticity}}}\ (\bibinfo  {publisher} {Butterworth-Heinemann},\ \bibinfo
  {address} {Oxford England Burlington, MA},\ \bibinfo {year}
  {1986})\BibitemShut {NoStop}%
\bibitem [{\citenamefont {Hutson}\ and\ \citenamefont
  {White}(1962)}]{Hutson1962}%
  \BibitemOpen
  \bibfield  {author} {\bibinfo {author} {\bibfnamefont {A.~R.}\ \bibnamefont
  {Hutson}}\ and\ \bibinfo {author} {\bibfnamefont {D.~L.}\ \bibnamefont
  {White}},\ }\href {\doibase 10.1063/1.1728525} {\bibfield  {journal}
  {\bibinfo  {journal} {J. Appl. Phys.}\ }\textbf {\bibinfo {volume} {33}},\
  \bibinfo {pages} {40} (\bibinfo {year} {1962})}\BibitemShut {NoStop}%
\bibitem [{\citenamefont {Morgan}(2010)}]{Morgan2010}%
  \BibitemOpen
  \bibfield  {author} {\bibinfo {author} {\bibfnamefont {D.}~\bibnamefont
  {Morgan}},\ }\href
  {https://books.google.com/books?hl=en\&lr=\&id=ITLn6tj4SDQC\&pgis=1} {\emph
  {\bibinfo {title} {Surface Acoustic Wave Filters: With Applications to
  Electronic Communications and Signal Processing}}}\ (\bibinfo  {publisher}
  {Academic Press},\ \bibinfo {address} {Amsterdam London},\ \bibinfo {year}
  {2010})\BibitemShut {NoStop}%
\bibitem [{\citenamefont {Auld}(1990)}]{Auld1990}%
  \BibitemOpen
  \bibfield  {author} {\bibinfo {author} {\bibfnamefont {B.}~\bibnamefont
  {Auld}},\ }\href {https://books.google.es/books?id=qUGEPwAACAAJ} {\emph
  {\bibinfo {title} {Acoustic Fields and Waves in Solids}}}\ (\bibinfo
  {publisher} {Krieger Publishing Company},\ \bibinfo {address} {Malabar,
  Fla},\ \bibinfo {year} {1990})\BibitemShut {NoStop}%
\bibitem [{\citenamefont {Royer}\ and\ \citenamefont
  {Dieulesaint}(2000{\natexlab{a}})}]{Royer2000}%
  \BibitemOpen
  \bibfield  {author} {\bibinfo {author} {\bibfnamefont {D.}~\bibnamefont
  {Royer}}\ and\ \bibinfo {author} {\bibfnamefont {E.}~\bibnamefont
  {Dieulesaint}},\ }\href
  {https://books.google.com/books?id=SzwQ1UYspyQC\&pgis=1} {\emph {\bibinfo
  {title} {Elastic Waves in Solids I: Free and Guided Propagation}}}\ (\bibinfo
   {publisher} {Springer Science \& Business Media},\ \bibinfo {address}
  {Berlin New York},\ \bibinfo {year} {2000})\BibitemShut {NoStop}%
\bibitem [{\citenamefont {Royer}\ and\ \citenamefont
  {Dieulesaint}(2000{\natexlab{b}})}]{Royer2000a}%
  \BibitemOpen
  \bibfield  {author} {\bibinfo {author} {\bibfnamefont {D.}~\bibnamefont
  {Royer}}\ and\ \bibinfo {author} {\bibfnamefont {E.}~\bibnamefont
  {Dieulesaint}},\ }\href
  {https://books.google.com/books?hl=en\&lr=\&id=q\_9rJYpgjZ0C\&pgis=1} {\emph
  {\bibinfo {title} {Elastic Waves in Solids II: Generation, Acousto-optic
  Interaction, Applications}}}\ (\bibinfo  {publisher} {Springer Science \&
  Business Media},\ \bibinfo {address} {Berlin New York},\ \bibinfo {year}
  {2000})\BibitemShut {NoStop}%
\bibitem [{\citenamefont {Newnham}(2004)}]{Newnham2004}%
  \BibitemOpen
  \bibfield  {author} {\bibinfo {author} {\bibfnamefont {R.~E.}\ \bibnamefont
  {Newnham}},\ }\href {https://books.google.com/books?id=YPyJVEXA-R8C\&pgis=1}
  {\emph {\bibinfo {title} {Properties of Materials : Anisotropy, Symmetry,
  Structure: Anisotropy, Symmetry, Structure}}},\ Vol.~\bibinfo {volume} {11}\
  (\bibinfo  {publisher} {OUP Oxford},\ \bibinfo {year} {2004})\BibitemShut
  {NoStop}%
\bibitem [{\citenamefont {Wixforth}\ \emph {et~al.}(1986)\citenamefont
  {Wixforth}, \citenamefont {Kotthaus},\ and\ \citenamefont
  {Weimann}}]{Wixforth1986}%
  \BibitemOpen
  \bibfield  {author} {\bibinfo {author} {\bibfnamefont {A.}~\bibnamefont
  {Wixforth}}, \bibinfo {author} {\bibfnamefont {J.~P.}\ \bibnamefont
  {Kotthaus}}, \ and\ \bibinfo {author} {\bibfnamefont {G.}~\bibnamefont
  {Weimann}},\ }\href@noop {} {\bibfield  {journal} {\bibinfo  {journal} {Phys.
  Rev. Lett.}\ }\textbf {\bibinfo {volume} {56}},\ \bibinfo {pages} {2104}
  (\bibinfo {year} {1986})}\BibitemShut {NoStop}%
\bibitem [{\citenamefont {{Castro Neto}}\ \emph {et~al.}(2009)\citenamefont
  {{Castro Neto}}, \citenamefont {Guinea}, \citenamefont {Peres}, \citenamefont
  {Novoselov},\ and\ \citenamefont {Geim}}]{CastroNeto2009}%
  \BibitemOpen
  \bibfield  {author} {\bibinfo {author} {\bibfnamefont {A.~H.}\ \bibnamefont
  {{Castro Neto}}}, \bibinfo {author} {\bibfnamefont {F.}~\bibnamefont
  {Guinea}}, \bibinfo {author} {\bibfnamefont {N.~M.~R.}\ \bibnamefont
  {Peres}}, \bibinfo {author} {\bibfnamefont {K.~S.}\ \bibnamefont
  {Novoselov}}, \ and\ \bibinfo {author} {\bibfnamefont {A.~K.}\ \bibnamefont
  {Geim}},\ }\href {\doibase 10.1103/RevModPhys.81.109} {\bibfield  {journal}
  {\bibinfo  {journal} {Rev. Mod. Phys.}\ }\textbf {\bibinfo {volume} {81}},\
  \bibinfo {pages} {109} (\bibinfo {year} {2009})}\BibitemShut {NoStop}%
\bibitem [{\citenamefont {Thalmeier}\ \emph {et~al.}(2010)\citenamefont
  {Thalmeier}, \citenamefont {D\'{o}ra},\ and\ \citenamefont
  {Ziegler}}]{Thalmeier2010}%
  \BibitemOpen
  \bibfield  {author} {\bibinfo {author} {\bibfnamefont {P.}~\bibnamefont
  {Thalmeier}}, \bibinfo {author} {\bibfnamefont {B.}~\bibnamefont {D\'{o}ra}},
  \ and\ \bibinfo {author} {\bibfnamefont {K.}~\bibnamefont {Ziegler}},\ }\href
  {\doibase 10.1103/PhysRevB.81.041409} {\bibfield  {journal} {\bibinfo
  {journal} {Phys. Rev. B}\ }\textbf {\bibinfo {volume} {81}},\ \bibinfo
  {pages} {041409} (\bibinfo {year} {2010})}\BibitemShut {NoStop}%
\bibitem [{\citenamefont {Miseikis}\ \emph {et~al.}(2012)\citenamefont
  {Miseikis}, \citenamefont {Cunningham}, \citenamefont {Saeed}, \citenamefont
  {O’Rorke},\ and\ \citenamefont {Davies}}]{Miseikis2012}%
  \BibitemOpen
  \bibfield  {author} {\bibinfo {author} {\bibfnamefont {V.}~\bibnamefont
  {Miseikis}}, \bibinfo {author} {\bibfnamefont {J.~E.}\ \bibnamefont
  {Cunningham}}, \bibinfo {author} {\bibfnamefont {K.}~\bibnamefont {Saeed}},
  \bibinfo {author} {\bibfnamefont {R.}~\bibnamefont {O’Rorke}}, \ and\
  \bibinfo {author} {\bibfnamefont {A.~G.}\ \bibnamefont {Davies}},\ }\href
  {\doibase 10.1063/1.3697403} {\bibfield  {journal} {\bibinfo  {journal}
  {Appl. Phys. Lett.}\ }\textbf {\bibinfo {volume} {100}},\ \bibinfo {pages}
  {133105} (\bibinfo {year} {2012})}\BibitemShut {NoStop}%
\bibitem [{\citenamefont {Bandhu}\ \emph {et~al.}(2013)\citenamefont {Bandhu},
  \citenamefont {Lawton},\ and\ \citenamefont {Nash}}]{Bandhu2013}%
  \BibitemOpen
  \bibfield  {author} {\bibinfo {author} {\bibfnamefont {L.}~\bibnamefont
  {Bandhu}}, \bibinfo {author} {\bibfnamefont {L.~M.}\ \bibnamefont {Lawton}},
  \ and\ \bibinfo {author} {\bibfnamefont {G.~R.}\ \bibnamefont {Nash}},\
  }\href {\doibase 10.1063/1.4822121} {\bibfield  {journal} {\bibinfo
  {journal} {Appl. Phys. Lett.}\ }\textbf {\bibinfo {volume} {103}},\ \bibinfo
  {pages} {133101} (\bibinfo {year} {2013})}\BibitemShut {NoStop}%
\bibitem [{\citenamefont {Zhang}\ \emph {et~al.}(2013)\citenamefont {Zhang},
  \citenamefont {Xu}, \citenamefont {Badalyan},\ and\ \citenamefont
  {Peeters}}]{Zhang2013}%
  \BibitemOpen
  \bibfield  {author} {\bibinfo {author} {\bibfnamefont {S.~H.}\ \bibnamefont
  {Zhang}}, \bibinfo {author} {\bibfnamefont {W.}~\bibnamefont {Xu}}, \bibinfo
  {author} {\bibfnamefont {S.~M.}\ \bibnamefont {Badalyan}}, \ and\ \bibinfo
  {author} {\bibfnamefont {F.~M.}\ \bibnamefont {Peeters}},\ }\href {\doibase
  10.1103/PhysRevB.87.075443} {\bibfield  {journal} {\bibinfo  {journal} {Phys.
  Rev. B}\ }\textbf {\bibinfo {volume} {87}},\ \bibinfo {pages} {075443}
  (\bibinfo {year} {2013})}\BibitemShut {NoStop}%
\bibitem [{\citenamefont {Schiefele}\ \emph {et~al.}(2013)\citenamefont
  {Schiefele}, \citenamefont {Pedr\'{o}s}, \citenamefont {Sols}, \citenamefont
  {Calle},\ and\ \citenamefont {Guinea}}]{Schiefele2013}%
  \BibitemOpen
  \bibfield  {author} {\bibinfo {author} {\bibfnamefont {J.}~\bibnamefont
  {Schiefele}}, \bibinfo {author} {\bibfnamefont {J.}~\bibnamefont
  {Pedr\'{o}s}}, \bibinfo {author} {\bibfnamefont {F.}~\bibnamefont {Sols}},
  \bibinfo {author} {\bibfnamefont {F.}~\bibnamefont {Calle}}, \ and\ \bibinfo
  {author} {\bibfnamefont {F.}~\bibnamefont {Guinea}},\ }\href {\doibase
  10.1103/PhysRevLett.111.237405} {\bibfield  {journal} {\bibinfo  {journal}
  {Phys. Rev. Lett.}\ }\textbf {\bibinfo {volume} {111}},\ \bibinfo {pages}
  {237405} (\bibinfo {year} {2013})}\BibitemShut {NoStop}%
\bibitem [{\citenamefont {Simon}(1996)}]{Simon1996}%
  \BibitemOpen
  \bibfield  {author} {\bibinfo {author} {\bibfnamefont {S.~H.}\ \bibnamefont
  {Simon}},\ }\href@noop {} {\bibfield  {journal} {\bibinfo  {journal} {Phys.
  Rev. B}\ }\textbf {\bibinfo {volume} {54}},\ \bibinfo {pages} {13878}
  (\bibinfo {year} {1996})}\BibitemShut {NoStop}%
\bibitem [{\citenamefont {Kn\"{a}bchen}\ \emph {et~al.}(1996)\citenamefont
  {Kn\"{a}bchen}, \citenamefont {Levinson},\ and\ \citenamefont
  {Entin-Wohlman}}]{Knabchen1996}%
  \BibitemOpen
  \bibfield  {author} {\bibinfo {author} {\bibfnamefont {A.}~\bibnamefont
  {Kn\"{a}bchen}}, \bibinfo {author} {\bibfnamefont {Y.~B.}\ \bibnamefont
  {Levinson}}, \ and\ \bibinfo {author} {\bibfnamefont {O.}~\bibnamefont
  {Entin-Wohlman}},\ }\href {\doibase 10.1103/PhysRevB.54.10696} {\bibfield
  {journal} {\bibinfo  {journal} {Phys. Rev. B}\ }\textbf {\bibinfo {volume}
  {54}},\ \bibinfo {pages} {10696} (\bibinfo {year} {1996})}\BibitemShut
  {NoStop}%
\bibitem [{\citenamefont {Farnell}(1970)}]{Farnell1970}%
  \BibitemOpen
  \bibfield  {author} {\bibinfo {author} {\bibfnamefont {G.~W.}\ \bibnamefont
  {Farnell}},\ }\href@noop {} {\bibfield  {journal} {\bibinfo  {journal}
  {Physical acoustics}\ }\textbf {\bibinfo {volume} {6}},\ \bibinfo {pages}
  {109} (\bibinfo {year} {1970})}\BibitemShut {NoStop}%
\bibitem [{\citenamefont {Mahan}(2013)}]{Mahan2013}%
  \BibitemOpen
  \bibfield  {author} {\bibinfo {author} {\bibfnamefont {G.~D.}\ \bibnamefont
  {Mahan}},\ }\href
  {https://books.google.com/books?hl=en\&lr=\&id=TFDUBwAAQBAJ\&pgis=1} {\emph
  {\bibinfo {title} {Many-Particle Physics}}}\ (\bibinfo  {publisher} {Springer
  Science \& Business Media},\ \bibinfo {address} {Boston, MA},\ \bibinfo
  {year} {2013})\BibitemShut {NoStop}%
\bibitem [{\citenamefont {Einenkel}\ and\ \citenamefont
  {Efetov}(2011)}]{Einenkel2011}%
  \BibitemOpen
  \bibfield  {author} {\bibinfo {author} {\bibfnamefont {M.}~\bibnamefont
  {Einenkel}}\ and\ \bibinfo {author} {\bibfnamefont {K.~B.}\ \bibnamefont
  {Efetov}},\ }\href {\doibase 10.1103/PhysRevB.84.214508} {\bibfield
  {journal} {\bibinfo  {journal} {Phys. Rev. B}\ }\textbf {\bibinfo {volume}
  {84}},\ \bibinfo {pages} {214508} (\bibinfo {year} {2011})}\BibitemShut
  {NoStop}%
\bibitem [{\citenamefont {Guinea}\ and\ \citenamefont
  {Uchoa}(2012)}]{Guinea2012}%
  \BibitemOpen
  \bibfield  {author} {\bibinfo {author} {\bibfnamefont {F.}~\bibnamefont
  {Guinea}}\ and\ \bibinfo {author} {\bibfnamefont {B.}~\bibnamefont {Uchoa}},\
  }\href {\doibase 10.1103/PhysRevB.86.134521} {\bibfield  {journal} {\bibinfo
  {journal} {Phys. Rev. B}\ }\textbf {\bibinfo {volume} {86}},\ \bibinfo
  {pages} {134521} (\bibinfo {year} {2012})}\BibitemShut {NoStop}%
\bibitem [{\citenamefont {Ingebrigsten}(1969)}]{Ingebrigsten1969}%
  \BibitemOpen
  \bibfield  {author} {\bibinfo {author} {\bibfnamefont {K.~A.}\ \bibnamefont
  {Ingebrigsten}},\ }\href@noop {} {\bibfield  {journal} {\bibinfo  {journal}
  {Journal of Applied Physics}\ }\textbf {\bibinfo {volume} {40}},\ \bibinfo
  {pages} {2681} (\bibinfo {year} {1969})}\BibitemShut {NoStop}%
\bibitem [{\citenamefont {Hwang}\ \emph {et~al.}(2010)\citenamefont {Hwang},
  \citenamefont {Sensarma},\ and\ \citenamefont {{Das Sarma}}}]{Hwang2010}%
  \BibitemOpen
  \bibfield  {author} {\bibinfo {author} {\bibfnamefont {E.~H.}\ \bibnamefont
  {Hwang}}, \bibinfo {author} {\bibfnamefont {R.}~\bibnamefont {Sensarma}}, \
  and\ \bibinfo {author} {\bibfnamefont {S.}~\bibnamefont {{Das Sarma}}},\
  }\href {\doibase 10.1103/PhysRevB.82.195406} {\bibfield  {journal} {\bibinfo
  {journal} {Phys. Rev. B}\ }\textbf {\bibinfo {volume} {82}},\ \bibinfo
  {pages} {195406} (\bibinfo {year} {2010})}\BibitemShut {NoStop}%
\bibitem [{\citenamefont {Wunsch}\ \emph {et~al.}(2006)\citenamefont {Wunsch},
  \citenamefont {Stauber}, \citenamefont {Sols},\ and\ \citenamefont
  {Guinea}}]{Wunsch2006}%
  \BibitemOpen
  \bibfield  {author} {\bibinfo {author} {\bibfnamefont {B.}~\bibnamefont
  {Wunsch}}, \bibinfo {author} {\bibfnamefont {T.}~\bibnamefont {Stauber}},
  \bibinfo {author} {\bibfnamefont {F.}~\bibnamefont {Sols}}, \ and\ \bibinfo
  {author} {\bibfnamefont {F.}~\bibnamefont {Guinea}},\ }\href {\doibase
  10.1088/1367-2630/8/12/318} {\bibfield  {journal} {\bibinfo  {journal} {New
  J. Phys.}\ }\textbf {\bibinfo {volume} {8}},\ \bibinfo {pages} {318}
  (\bibinfo {year} {2006})}\BibitemShut {NoStop}%
\bibitem [{\citenamefont {Giuliani}\ and\ \citenamefont
  {Vignale}(2005)}]{Giuliani2005}%
  \BibitemOpen
  \bibfield  {author} {\bibinfo {author} {\bibfnamefont {G.}~\bibnamefont
  {Giuliani}}\ and\ \bibinfo {author} {\bibfnamefont {G.}~\bibnamefont
  {Vignale}},\ }\href {https://books.google.com/books?id=kFkIKRfgUpsC\&pgis=1}
  {\emph {\bibinfo {title} {Quantum Theory of the Electron Liquid}}}\ (\bibinfo
   {publisher} {Cambridge University Press},\ \bibinfo {address} {Cambridge},\
  \bibinfo {year} {2005})\BibitemShut {NoStop}%
\bibitem [{\citenamefont {Mattuck}(2012)}]{Mattuck2012}%
  \BibitemOpen
  \bibfield  {author} {\bibinfo {author} {\bibfnamefont {R.~D.}\ \bibnamefont
  {Mattuck}},\ }\href {https://books.google.com/books?id=1P\_DAgAAQBAJ\&pgis=1}
  {\emph {\bibinfo {title} {A Guide to Feynman Diagrams in the Many-Body
  Problem}}}\ (\bibinfo  {publisher} {Courier Corporation},\ \bibinfo {address}
  {New York},\ \bibinfo {year} {2012})\BibitemShut {NoStop}%
\bibitem [{\citenamefont {Ashcroft}\ and\ \citenamefont
  {Mermin}(2011)}]{Ashcroft2011}%
  \BibitemOpen
  \bibfield  {author} {\bibinfo {author} {\bibfnamefont {N.~W.}\ \bibnamefont
  {Ashcroft}}\ and\ \bibinfo {author} {\bibfnamefont {N.~D.}\ \bibnamefont
  {Mermin}},\ }\href {https://books.google.com/books?id=x\_s\_YAAACAAJ\&pgis=1}
  {\emph {\bibinfo {title} {Solid State Physics}}}\ (\bibinfo  {publisher}
  {Cengage Learning},\ \bibinfo {address} {New York},\ \bibinfo {year}
  {2011})\BibitemShut {NoStop}%
\bibitem [{\citenamefont {Gor'kov}(2016)}]{Gorkov2015}%
  \BibitemOpen
  \bibfield  {author} {\bibinfo {author} {\bibfnamefont {L.~P.}\ \bibnamefont
  {Gor'kov}},\ }\href {\doibase 10.1103/PhysRevB.93.060507} {\bibfield
  {journal} {\bibinfo  {journal} {Phys. Rev. B}\ }\textbf {\bibinfo {volume}
  {93}},\ \bibinfo {pages} {060507} (\bibinfo {year} {2016})}\BibitemShut
  {NoStop}%
\bibitem [{\citenamefont {Kohn}\ and\ \citenamefont
  {Luttinger}(1965)}]{kohn1965new}%
  \BibitemOpen
  \bibfield  {author} {\bibinfo {author} {\bibfnamefont {W.}~\bibnamefont
  {Kohn}}\ and\ \bibinfo {author} {\bibfnamefont {J.}~\bibnamefont
  {Luttinger}},\ }\href@noop {} {\bibfield  {journal} {\bibinfo  {journal}
  {Physical Review Letters}\ }\textbf {\bibinfo {volume} {15}},\ \bibinfo
  {pages} {524} (\bibinfo {year} {1965})}\BibitemShut {NoStop}%
\bibitem [{\citenamefont {Morel}\ and\ \citenamefont
  {Anderson}(1962)}]{Morel1962}%
  \BibitemOpen
  \bibfield  {author} {\bibinfo {author} {\bibfnamefont {P.}~\bibnamefont
  {Morel}}\ and\ \bibinfo {author} {\bibfnamefont {P.~W.}\ \bibnamefont
  {Anderson}},\ }\href {\doibase 10.1103/PhysRev.125.1263} {\bibfield
  {journal} {\bibinfo  {journal} {Phys. Rev.}\ }\textbf {\bibinfo {volume}
  {125}},\ \bibinfo {pages} {1263} (\bibinfo {year} {1962})}\BibitemShut
  {NoStop}%
\bibitem [{\citenamefont {Gonz\'{a}lez}\ \emph {et~al.}(shed)\citenamefont
  {Gonz\'{a}lez}, \citenamefont {Schiefele}, \citenamefont {Sols},
  \citenamefont {Guinea},\ and\ \citenamefont {Zapata}}]{Gonzalez2015}%
  \BibitemOpen
  \bibfield  {author} {\bibinfo {author} {\bibfnamefont {D.~G.}\ \bibnamefont
  {Gonz\'{a}lez}}, \bibinfo {author} {\bibfnamefont {J.}~\bibnamefont
  {Schiefele}}, \bibinfo {author} {\bibfnamefont {F.}~\bibnamefont {Sols}},
  \bibinfo {author} {\bibfnamefont {F.}~\bibnamefont {Guinea}}, \ and\ \bibinfo
  {author} {\bibfnamefont {I.}~\bibnamefont {Zapata}},\ }\href@noop {} {\
  (\bibinfo {year} {unpublished})}\BibitemShut {NoStop}%
\bibitem [{Note1()}]{Note1}%
  \BibitemOpen
  \bibinfo {note} {In this analysis, we are taking into account only
  long-wavelength piezoelectric phonons, hence no intervalley pairing
  instability could occur. We address this question in the next
  paragraph.}\BibitemShut {Stop}%
\bibitem [{\citenamefont {Ye}\ \emph {et~al.}(2012)\citenamefont {Ye},
  \citenamefont {Zhang}, \citenamefont {Akashi}, \citenamefont {Bahramy},
  \citenamefont {Arita},\ and\ \citenamefont {Iwasa}}]{Ye2012}%
  \BibitemOpen
  \bibfield  {author} {\bibinfo {author} {\bibfnamefont {J.~T.}\ \bibnamefont
  {Ye}}, \bibinfo {author} {\bibfnamefont {Y.~J.}\ \bibnamefont {Zhang}},
  \bibinfo {author} {\bibfnamefont {R.}~\bibnamefont {Akashi}}, \bibinfo
  {author} {\bibfnamefont {M.~S.}\ \bibnamefont {Bahramy}}, \bibinfo {author}
  {\bibfnamefont {R.}~\bibnamefont {Arita}}, \ and\ \bibinfo {author}
  {\bibfnamefont {Y.}~\bibnamefont {Iwasa}},\ }\href {\doibase
  10.1126/science.1228006} {\bibfield  {journal} {\bibinfo  {journal} {Science
  (New York, N.Y.)}\ }\textbf {\bibinfo {volume} {338}},\ \bibinfo {pages}
  {1193} (\bibinfo {year} {2012})}\BibitemShut {NoStop}%
\bibitem [{\citenamefont {Rold\'{a}n}\ \emph {et~al.}(2013)\citenamefont
  {Rold\'{a}n}, \citenamefont {Cappelluti},\ and\ \citenamefont
  {Guinea}}]{Roldan2013}%
  \BibitemOpen
  \bibfield  {author} {\bibinfo {author} {\bibfnamefont {R.}~\bibnamefont
  {Rold\'{a}n}}, \bibinfo {author} {\bibfnamefont {E.}~\bibnamefont
  {Cappelluti}}, \ and\ \bibinfo {author} {\bibfnamefont {F.}~\bibnamefont
  {Guinea}},\ }\href {\doibase 10.1103/PhysRevB.88.054515} {\bibfield
  {journal} {\bibinfo  {journal} {Phys. Rev. B}\ }\textbf {\bibinfo {volume}
  {88}},\ \bibinfo {pages} {054515} (\bibinfo {year} {2013})}\BibitemShut
  {NoStop}%
\bibitem [{\citenamefont {Hashimoto}(2000)}]{Hashimoto2000}%
  \BibitemOpen
  \bibfield  {author} {\bibinfo {author} {\bibfnamefont {K.-y.}\ \bibnamefont
  {Hashimoto}},\ }\href@noop {} {\emph {\bibinfo {title} {Surface acoustic wave
  devices in telecommunications}}}\ (\bibinfo  {publisher} {Springer},\
  \bibinfo {address} {Berlin},\ \bibinfo {year} {2000})\BibitemShut {NoStop}%
\bibitem [{\citenamefont {Lothe}\ and\ \citenamefont
  {Barnett}(1976)}]{Lothe1976}%
  \BibitemOpen
  \bibfield  {author} {\bibinfo {author} {\bibfnamefont {J.}~\bibnamefont
  {Lothe}}\ and\ \bibinfo {author} {\bibfnamefont {D.~M.}\ \bibnamefont
  {Barnett}},\ }\href {\doibase 10.1063/1.322895} {\bibfield  {journal}
  {\bibinfo  {journal} {J. Appl. Phys.}\ }\textbf {\bibinfo {volume} {47}},\
  \bibinfo {pages} {1799} (\bibinfo {year} {1976})}\BibitemShut {NoStop}%
\bibitem [{\citenamefont {Tiersten}(1969)}]{Tiersten1969}%
  \BibitemOpen
  \bibfield  {author} {\bibinfo {author} {\bibfnamefont {H.~F.}\ \bibnamefont
  {Tiersten}},\ }\href@noop {} {\emph {\bibinfo {title} {Linear Piezoelectric
  Plate Vibrations}}}\ (\bibinfo  {publisher} {Plenum Press},\ \bibinfo
  {address} {New York},\ \bibinfo {year} {1969})\BibitemShut {NoStop}%
\bibitem [{\citenamefont {Kubo}\ \emph {et~al.}(2012)\citenamefont {Kubo},
  \citenamefont {Toda},\ and\ \citenamefont {Hashitsume}}]{Kubo2012}%
  \BibitemOpen
  \bibfield  {author} {\bibinfo {author} {\bibfnamefont {R.}~\bibnamefont
  {Kubo}}, \bibinfo {author} {\bibfnamefont {M.}~\bibnamefont {Toda}}, \ and\
  \bibinfo {author} {\bibfnamefont {N.}~\bibnamefont {Hashitsume}},\ }\href
  {https://books.google.es/books?id=Z0DynQEACAAJ} {\emph {\bibinfo {title}
  {Statistical Physics II: Nonequilibrium Statistical Mechanics}}},\ Springer
  series in solid-state sciences\ (\bibinfo  {publisher} {Springer},\ \bibinfo
  {address} {Berlin},\ \bibinfo {year} {2012})\BibitemShut {NoStop}%
\bibitem [{\citenamefont {Darinskii}\ \emph {et~al.}(2007)\citenamefont
  {Darinskii}, \citenamefont {{Le Clezio}},\ and\ \citenamefont
  {Feuillard}}]{Darinskii2007}%
  \BibitemOpen
  \bibfield  {author} {\bibinfo {author} {\bibfnamefont {A.~N.}\ \bibnamefont
  {Darinskii}}, \bibinfo {author} {\bibfnamefont {E.}~\bibnamefont {{Le
  Clezio}}}, \ and\ \bibinfo {author} {\bibfnamefont {G.}~\bibnamefont
  {Feuillard}},\ }\href {\doibase 10.1109/TUFFC.2007.284} {\bibfield  {journal}
  {\bibinfo  {journal} {IEEE Transactions on Ultrasonics, Ferroelectrics, and
  Frequency Control}\ }\textbf {\bibinfo {volume} {54}},\ \bibinfo {pages}
  {612} (\bibinfo {year} {2007})}\BibitemShut {NoStop}%
\end{thebibliography}%

\end{document}